\documentclass{article} 

\usepackage{iclr2027_conference,times}

\usepackage{amsmath,amsfonts,bm}

\def\eqref#1{equation~\ref{#1}}

\def\1{\bm{1}}

\DeclareMathAlphabet{\mathsfit}{\encodingdefault}{\sfdefault}{m}{sl}
\SetMathAlphabet{\mathsfit}{bold}{\encodingdefault}{\sfdefault}{bx}{n}

\usepackage{microtype}
\usepackage{graphicx}
\usepackage{xspace}
\usepackage{subcaption}
\usepackage[normalem]{ulem}
\usepackage{booktabs}
\usepackage{hyperref}
\usepackage{url}
\usepackage{xurl}
\usepackage{pifont}
\usepackage{soul}
\usepackage{amsmath}
\usepackage{amssymb}
\usepackage{mathtools}
\usepackage{amsthm}
\usepackage[capitalize,noabbrev]{cleveref}
\usepackage[boxed, ruled, vlined]{algorithm2e}
\usepackage{algorithmic}
\usepackage{multirow}
\usepackage{colortbl}
\usepackage{xcolor}
\usepackage[most]{tcolorbox}
\usepackage{array}
\usepackage{wrapfig}

\definecolor{cellcolor}{RGB}{0,0,0}
\definecolor{antiquewhite}{rgb}{0.98, 0.92, 0.84}
\definecolor{anti-flashwhite}{rgb}{0.95, 0.95, 0.96}
\definecolor{aliceblue}{rgb}{0.94, 0.97, 1.0}
\definecolor{almond}{rgb}{0.94, 0.87, 0.8}
\definecolor{cosmiclatte}{rgb}{1.0, 0.97, 0.91}
\definecolor{darkbyzantium}{rgb}{0.36, 0.22, 0.33}
\definecolor{darkseagreen}{rgb}{0.56, 0.74, 0.56}
\definecolor{darkspringgreen}{rgb}{0.09, 0.45, 0.27}
\definecolor{asparagus}{rgb}{0.53, 0.66, 0.42}
\definecolor{antiquefuchsia}{rgb}{0.57, 0.36, 0.51}
\definecolor{ao(english)}{rgb}{0.0, 0.5, 0.0}
\definecolor{deepcerise}{rgb}{0.85, 0.2, 0.53}
\definecolor{denim}{rgb}{0.08, 0.38, 0.74}
\definecolor{crimson}{rgb}{0.86, 0.08, 0.24}
\definecolor{buff}{rgb}{0.94, 0.86, 0.51}
\definecolor{amber(sae/ece)}{rgb}{1.0, 0.49, 0.0}
\definecolor{airforceblue}{rgb}{0.36, 0.54, 0.66}
\definecolor{amethyst}{rgb}{0.6, 0.4, 0.8}
\definecolor{azure(colorwheel)}{rgb}{0.0, 0.5, 1.0}
\definecolor{azure(web)(azuremist)}{rgb}{0.94, 1.0, 1.0}
\definecolor{beige}{rgb}{0.96, 0.96, 0.86}
\definecolor{cornsilk}{rgb}{1.0, 0.97, 0.86}
\definecolor{darkcerulean}{rgb}{0.03, 0.27, 0.49}
\definecolor{babyblue}{rgb}{0.54, 0.81, 0.94}
\definecolor{ballblue}{rgb}{0.13, 0.67, 0.8}
\definecolor{dukeblue}{rgb}{0.0, 0.0, 0.61}
\definecolor{champagne}{rgb}{0.97, 0.91, 0.81}
\definecolor{antiquefuchsia}{rgb}{0.57, 0.36, 0.51}
\definecolor{darkgreen}{rgb}{0, 0.5, 0}

\definecolor{cellshade}{HTML}{F0EDFF}

\usepackage{wrapfig}

\newtcolorbox{algobox}[1]{
    enhanced,
    colback=blue!2,
    colframe=dukeblue,
    coltitle=black,
    fonttitle=\bfseries\small,
    title={#1},
    boxrule=0.8pt,
    arc=2pt,
    left=4pt,
    right=4pt,
    top=4pt,
    bottom=4pt,
    fontupper=\footnotesize,
    before skip=0pt,
    after skip=0pt
}

\SetAlCapNameFnt{\scriptsize\bfseries}
\SetAlCapFnt{\scriptsize}

\theoremstyle{plain}

\theoremstyle{definition}

\theoremstyle{remark}

\definecolor{headgray}{RGB}{245,245,248}
\definecolor{groupA}{RGB}{232,246,255}
\definecolor{groupB}{RGB}{243,232,255}
\definecolor{groupC}{RGB}{255,240,245}
\definecolor{row1}{RGB}{255,255,255}
\definecolor{row2}{RGB}{250,250,250}
\definecolor{goodgreen}{RGB}{0,128,0}
\definecolor{badred}{RGB}{200,0,0}

\newcommand{\gooddown}[1]{{\color{goodgreen}$\downarrow$} #1}

\newcommand{\goodup}[1]{{\color{goodgreen}$\uparrow$}\xspace#1}
\newcommand{\baddown}[1]{{\color{badred}$\downarrow$}\xspace#1}

\newcommand{\system}{\textsc{Polar}\xspace}

\newcommand{\mycode}{
\url{https://github.com/ltang24/LLM-Prioritization}\xspace
}

\definecolor{almond}{rgb}{0.94, 0.87, 0.8}
\definecolor{palesilver}{rgb}{0.79, 0.75, 0.73}

\tcbset{
  colback=palesilver!20,
  colframe=black!80,
  coltitle=black!70,
  boxrule=0.8pt,
  arc=4pt,
  left=3pt,
  right=3pt,
  top=3pt,
  bottom=3pt,
  fonttitle=\bfseries\large\color{white}
}

\newtcolorbox{bluecolorbox}{
  colback=blue!3,
  colframe=blue!75!black,
  coltitle=black,
  fonttitle=\bfseries,
  boxrule=1.5pt,
  arc=3pt,
  left=2pt,
  right=2pt,
  top=1pt,
  bottom=1pt
}

\newtcolorbox{casebox}[2][]{
    enhanced,
    breakable,
    colframe=asparagus,
    colback=white,
    coltitle=black,
    fonttitle=\bfseries\small,
    fontupper=\small,
    title={#2},
    attach boxed title to top left={
        xshift=2mm,
        yshift*=-3mm
    },
    boxed title style={
        colback=asparagus!10,
        sharp corners,
        frame hidden,
    },
    sharp corners,
    boxrule=0.4mm,
    top=-2.5pt,
    bottom=1.5pt,
    left=2.5pt,
    right=2.5pt,
    width=\linewidth,
    #1
}

\tcbset{
  polarbox/.style={
    breakable,
    enhanced,
    colback=white,
    colframe=black!35,
    boxrule=0.4pt,
    sharp corners,
    arc=2pt,
    left=6pt,
    right=6pt,
    top=6pt,
    bottom=6pt,
    boxsep=3pt,
    titlerule=0pt,
    title style={font=\bfseries},
    before skip=6pt,
    after skip=8pt
  }
}

\newtcolorbox{promptbox}[1]{
    polarbox,
    title={#1}
}

\title{
Polar: LLM-Powered Synthesis of Real-World Cyber Evidence
for Prioritization and Mitigation
}

\author{
\textbf{Luoxi Tang}$^{1}$ \;
\textbf{Yuqiao Meng}$^{1}$ \;
\textbf{Ankita Patra}$^{1}$ \;
\textbf{Weicheng Ma}$^{2}$ \;
\textbf{Muchao Ye}$^{3}$ \; \textbf{Zhaohan Xi}$^{1}$  \\[2mm]
$^{1}$Binghamton University, State University of New York,
Binghamton, NY, USA \\
$^{2}$Oakland University,
Rochester, MI, USA \\
$^{3}$University of Iowa,
Iowa City, IA, USA \\[1mm]
\texttt{\{ltang24,ymeng15,zxi1\}@binghamton.edu}
}

\iclrfinalcopy

\AddToHook{cmd/maketitle/after}{
    \fancyhf{}
    \fancyhead[L]{Preprint.}
    \renewcommand{\headrulewidth}{0.4pt}
    \pagestyle{fancy}
    \thispagestyle{fancy}
}

\begin{document}

\maketitle


\begin{abstract}
    Cyber threat analysis increasingly depends on evidence distributed across vendor advisories, vulnerability databases, and threat intelligence sources. Turning these fragmented observations into timely decisions requires models to connect technical severity with evolving exploitation evidence and available defensive actions. We present \system, an LLM-powered framework for synthesizing real-world cyber evidence into threat-centric assessments for prioritization and mitigation. \system first disentangles overlapping incidents and grounds each threat in source-linked evidence. For prioritization, it infers severity metrics from cyber evidence and combines the resulting assessment with temporally ordered exploitation signals to estimate near-term exploitation likelihood. For mitigation, it links the synthesized threat data to authoritative remediation knowledge and organizes applicable actions according to threat urgency and operational constraints. We evaluate \system on real-world vulnerability evidence collected from public resources and compare it with multiple baselines. Across heterogeneous incidents and zero-day settings, \system improves threat ranking and mitigation retrieval while producing evidence-linked intermediate assessments that support analyst inspection. The results establish evidence synthesis as a practical foundation for LLM-based cyber decision support across related security tasks.
\end{abstract}

\section{Introduction}

Cyber threat analysis increasingly depends on evidence that is fragmented across real-world security sources. A vulnerability may first appear in a vendor disclosure, later receive structured metadata in a vulnerability database, and then acquire new operational significance as exploitation evidence emerges. We use the term \textbf{real-world cyber evidence} to describe these observations and authoritative records as they accumulate around a threat. The central analytical problem is to turn this evolving evidence into decisions about threat urgency and remediation. Standard mechanisms such as the Common Vulnerability Scoring System (CVSS) \citep{firstCVSS31spec,firstCVSS31userguide} and the Exploit Prediction Scoring System (EPSS) \citep{jacobs2021epss,Jacobs2019EPSSarxiv} provide useful views of severity and exploitation likelihood, yet threat analysis still requires evidence to be synthesized across sources and interpreted in context.

Existing cyber analysis systems capture only part of this process. Rule-based tools and platforms are effective when signatures and decision rules are known \citep{white2013quantitative,sancho2020new} . Their coverage weakens as threats and evidence evolve \citep{barr2021survivalism,dong2023distdet,liu2024survey}. Learning-based systems offer greater flexibility, including vulnerability classifiers \citep{almutairi2026llm,zhan2021atvhunter} and log-analysis models \citep{sasikrishna2026llm,pakshad2026llm}, though they commonly rely on fixed input representations and remain sensitive to shifts in CTI sources. A deeper limitation is that threat signals are often processed as isolated features. In practice, their meaning changes with context \citep{jorgensen2026cybersecurity}. A moderate-severity vulnerability can become urgent after credible exploitation appears, while a high-severity vulnerability may pose limited immediate risk under restrictive exploitation conditions \citep{eling2023time}. Reliable analysis therefore depends on connecting evidence across technical and temporal context.

\begin{figure}[t]
    \centering
    \includegraphics[width=0.98\linewidth]{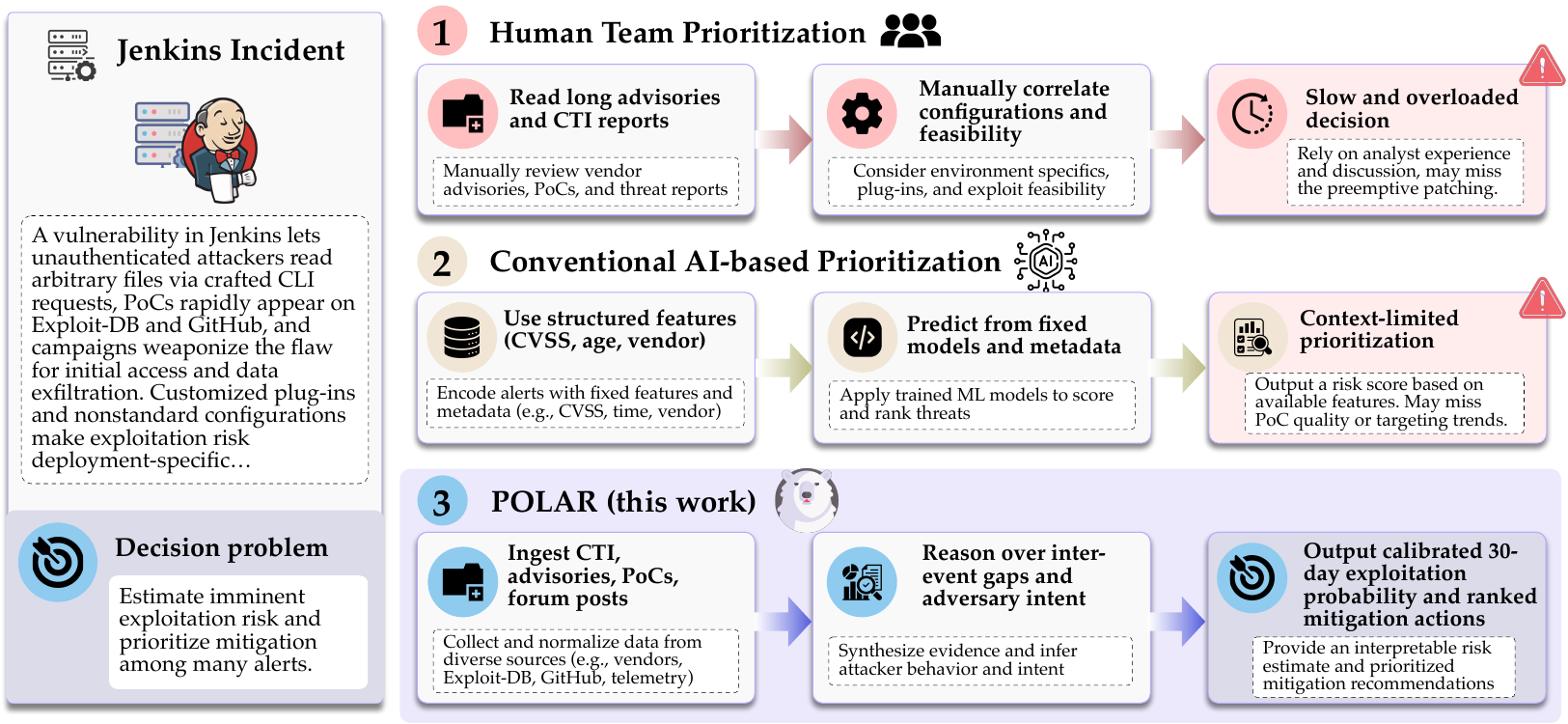}
    \caption{A motivating example of different threat-prioritization workflows, wherein \system synthesizes heterogeneous cyber evidence to support timely assessment and mitigation planning.}
    \label{fig:illustration}
\end{figure}

Large language models (LLMs) offer opportunities for effective threat analysis given their capability to synthesize evidence expressed at different levels of abstraction \citep{nguyen2026deepening}, such as an authentication requirement in an advisory and a subsequent report of active exploitation.  Recent work has applied LLMs to penetration testing \citep{deng2024pentestgpt}, attack-path analysis \citep{prapty2024using,bahlibi2026beyond,meng2025benchmarking}, and security-event reasoning \citep{uetz2024you,meng2025uncovering}. The remaining challenge is to organize this evidence before it is used for a decision. However, direct prompt-based agent design can conflate incidents or obscure the evidence that supports a conclusion \citep{ramachandran2026prompt}.
This leads to our research question: \ul{How can LLMs synthesize heterogeneous real-world cyber evidence for structured assessments and downstream decisions?} 

\textbf{This Work.} We present \system (Figure~\ref{fig:illustration}), an LLM-powered framework that builds a threat-centric evidence profile from unstructured incidents and authoritative external sources. \system separates overlapping threat descriptions, grounds each threat in source-linked metadata, and preserves evidence relevant to later analysis. This profile serves as the common input to two application scenarios studied in this work: \textbf{(I) Prioritization:} \system combines two complementary assessments. It derives CVSS metrics from evidence specific to each severity dimension, then applies the official scoring procedure to obtain a standardized severity estimate. It also organizes exploitation evidence by time and uses the resulting history to estimate near-term exploitation likelihood. The combined view captures the consequence of successful exploitation together with the current evidence that exploitation may occur. \textbf{(II) Mitigation:} \system connects the same threat profile to authoritative remediation knowledge and identifies actions applicable to the affected product and exploitation context. The resulting recommendations use the threat assessment and operational constraints to connect evidence of urgency with evidence for action.

We evaluate \system on real-world vulnerability data enriched with temporal threat events, covering a broad range of CVEs and cross-referencing evidence from Exploit-DB \citep{exploit-db}, CISA KEV \citep{CISA_KEV_Catalog}, VirusTotal \citep{virustotal}, and vendor advisories. Across comparisons with different baseline methods, \system consistently improves threat prioritization and remains stable under entangled and heterogeneous incident settings. The gains are significant for high-severity threats with active or rapidly evolving exploitation evidence, where \system produces more robust and timely prioritization. These results demonstrate the value of synthesizing heterogeneous real-world cyber evidence for decision support.

Overall, this work makes several contributions: (1) We formulate real-world cyber evidence synthesis as a unified foundation for LLM-based prioritization and mitigation. (2) We develop \system to organize heterogeneous cyber evidence into source-grounded threat assessments combining severity and exploitation dynamics. (3) We evaluate \system across prioritization and mitigation tasks, showing consistent gains over multiple baseline methods.

\section{Motivating Example}
\label{sec:example}

Figure~\ref{fig:illustration} uses a Jenkins vulnerability to illustrate the practical difficulty of threat prioritization. The vulnerability allows an unauthenticated attacker to read arbitrary files through crafted CLI requests. Soon after disclosure, public PoCs appear on Exploit-DB and GitHub, and attacker campaigns begin using the flaw for initial access and data exfiltration. The risk also varies across deployments because Jenkins instances may use different plug-ins and nonstandard configurations. A security team therefore needs to answer two questions quickly: \ul{How likely is this vulnerability to be exploited in the near term, and what should be mitigated first?}

\textbf{Human team prioritization:}
An analyst would first read the vendor advisory and CTI reports, then check the available PoCs and compare them with the local Jenkins configuration. The analyst may also need to determine whether the exposed CLI path is reachable, whether installed plug-ins change the attack surface, and whether the observed campaign applies to the local environment. This process can produce a careful assessment, but it takes time. When hundreds of alerts arrive together, analysts must repeat the same cross-checking for many threats while new exploitation evidence continues to appear. The resulting delay can consume the short window in which preemptive patching is most useful.

\textbf{Conventional AI-based prioritization:}
A conventional model can score the same vulnerability much faster from structured inputs such as CVSS metrics, vulnerability age, vendor information, and available enrichment metadata. This works well when the relevant signals are already encoded as features. The difficulty is that some of the most useful signals arrive as text or emerge after disclosure. The quality of a new PoC, evidence of active targeting, changes in attacker behavior, or deployment-specific constraints may never appear in the feature vector. Missing patch or exploit metadata can further weaken the score. The model can therefore rank the vulnerability quickly while still missing changes that make the threat more urgent in practice.

\textbf{\system prioritization:}
\system processes the same incident together with the surrounding cyber evidence. It collects the advisory, CTI reports, public PoCs, exploitation reports, and other source-linked observations, then organizes them into a temporal sequence such as
$\text{disclosure}\rightarrow\text{PoC}\rightarrow\text{KEV}\rightarrow\text{in-the-wild exploitation}$.
The model examines how quickly these events occur, whether the PoC is readily weaponizable, and whether current attacker activity matches the affected environment. From this evidence, \system estimates the probability of exploitation within the next 30 days and combines that assessment with the threat's technical severity. It then links the prioritized threat to applicable mitigation evidence, such as vendor patches, workarounds, and defensive controls.

This example illustrates the key design principle of \system: cyber prioritization  depends on evidence that changes over time and arrives in different forms. A useful system must keep those signals connected to the threat being assessed and carry them through to the mitigation decision.
\section{\system: Real-World Cyber Evidence Synthesis}
\label{sec:method}

\begin{wrapfigure}{r}{0.41\columnwidth}
\vspace{-1.5em}
\begin{minipage}{\linewidth}
\scriptsize
\begin{algorithm}[H]
  \SetAlgoLined
  \caption{Cyber Evidence Synthesis
  \label{algo:synthesis}}

  \KwIn{Incidents $\mathcal{I}=\{i_j\}_{j=1}^{N}$; evidence sources $\mathcal{D}$}
  \KwOut{Evidence $\mathcal{Z}=\{z_k\}_{k=1}^{M}$}

  \tcp{Disentangle threat instances}
  \ForEach{$i_j\in\mathcal{I}$}{
    $\mathcal{T}_j
    \leftarrow
    \phi_{\mathrm{LLM}}(i_j)
    =
    \{\tau_{j,1},\ldots,\tau_{j,L_j}\}$\;
  }

  \tcp{Ground threats in external evidence}
  \ForEach{$\tau_k\in\bigcup_j\mathcal{T}_j$}{
    $\mathcal{E}_k
    \leftarrow
    \psi(\tau_k,\mathcal{D})$\;
    $\mathbf{m}_k
    \leftarrow
    \textsc{Metadata}(\tau_k,\mathcal{D})$\;
    $z_k
    \leftarrow
    (\tau_k,\mathbf{m}_k,\mathcal{E}_k)$\;
  }

  \Return $\mathcal{Z}=\{z_k\}_{k=1}^{M}$\;
\end{algorithm}
\end{minipage}
\vspace{-3.5em}
\end{wrapfigure}

\begin{figure}[t]
    \centering
    \includegraphics[width=\linewidth]{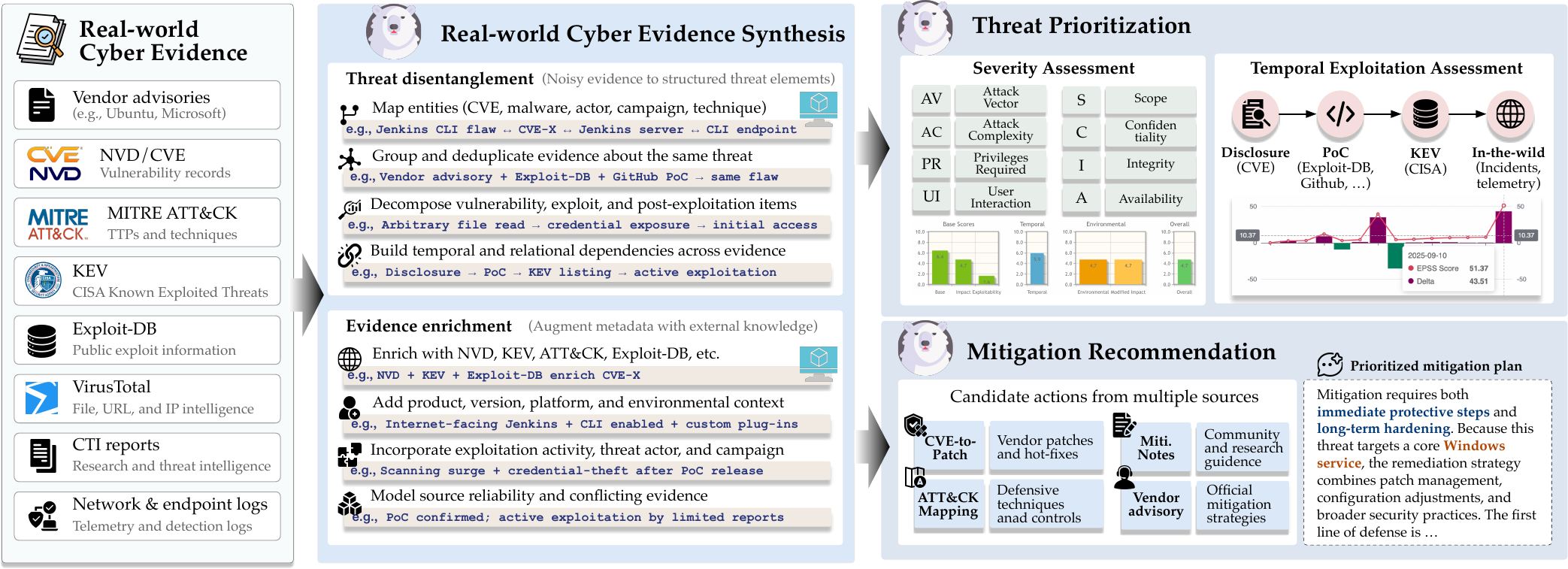}
    \caption{Overview of \system. Real-world cyber evidence is synthesized into structured threat profiles that support downstream threat prioritization and mitigation recommendation.}
    \label{fig:workflow}
\end{figure}

\system synthesizes heterogeneous cyber reports into threat-centric evidence for downstream security analysis. Figure~\ref{fig:workflow} illustrates the workflow. Below we detail the design of \system.



\subsection{Real-World Cyber Evidence Synthesis}
\label{ssec:evidence-synthesis}

Cyber evidence is distributed across reports that differ in scope, structure, and timing. A vendor advisory may describe several vulnerabilities together, while exploit repositories and later threat reports contribute new information about weaponization and deployment conditions. We formulate evidence synthesis as the construction of a structured representation (Algorithm \ref{algo:synthesis}) centered on an individual threat while retaining the provenance and temporal context of its supporting evidence.

\textbf{\ding{182} Threat disentanglement.}
Let $\mathcal{I}=\{i_1,\ldots,i_N\}$ denote the raw incidents available at analysis time $t$. An incident may describe several vulnerabilities, campaigns, or related indicators. \system applies an LLM-based separation operator $\phi_{\mathrm{LLM}}$ to obtain canonical threat instances $\tau_k$, while preserving the affected component, exploitation path, and deployment conditions associated with each instance. This separation reduces evidence conflation when multiple threats appear in the same report. Detailed instructions are provided in Appendix~\ref{app:evi-seynthesis}.

\textbf{\ding{183} Source-aware evidence enrichment.}
For each threat $\tau_k$, \system retrieves complementary evidence from external sources $\mathcal{D}$, including vulnerability databases, ATT\&CK knowledge, exploitation feeds, and vendor or government advisories. The resulting set
\smash{$\mathcal{E}_k=\psi(\tau_k,\mathcal{D})=\{e_{k,1},\ldots,e_{k,R_k}\}$}
contains evidence units
\smash{$e_{k,r}=(s_{k,r},u_{k,r},t_{k,r},c_{k,r})$},
where \smash{$s_{k,r}$} denotes the supporting content, \smash{$u_{k,r}$} as its source, \smash{$t_{k,r}$} as its timestamp when available, and $c_{k,r}$ as its analytical role. In parallel, \system extracts standardized metadata $\mathbf{m}_k$, including the CVE identifier, affected product, disclosure date, and ATT\&CK mappings. Together they form the threat-centric evidence
$z_k=(\tau_k,\mathbf{m}_k,\mathcal{E}_k)$.

Note that  $z_k$ includes severity and temporal evidence, thus facilitating subsequent CVSS assessment and exploitation analysis. Given an analytical query $q$ (i.e., a task), \system selects the corresponding view
\smash{$\mathcal{E}_k^{(q)}\subseteq\mathcal{E}_k$} for task-specific analysis
while retaining provenance and contextual dependencies.

\vspace{-3pt}
\begin{casebox}{\bf Example 1: Synthesizing evidence for a Print Spooler threat}
\ul{\bf Raw incident $i_j$:} A Windows 11 report states that an authenticated remote attacker can abuse the Print Spooler through a crafted DCERPC request to load a DLL and execute code as NT AUTHORITY.

\vspace{2pt}
\ul{\bf Threat instance $\tau_k$:} \system disentangles this report into a canonical Print Spooler threat, preserving its affected component, remote exploitation path, and privilege conditions.

\vspace{2pt}
\ul{\bf Evidence profile $z_k$:} \system extracts metadata $\mathbf{m}_k$ (e.g., the mapped CVE, affected versions, and ATT\&CK techniques), and constructs $\mathcal{E}_k$ from source-linked evidence units $e_{k,r}=(s_{k,r},u_{k,r},t_{k,r},c_{k,r})$. For example, ``authenticated remote attacker'' is retained as severity evidence, while a CISA KEV record contributes temporally grounded exploitation evidence. 
\end{casebox}

\vspace{-3pt}
Applying this procedure to all incidents yields
$\mathcal{Z}=\{z_1,\ldots,z_M\}$,
which is shared by the two applications below.

\subsection{Application I: Cyber Threat Prioritization}
\label{ssec:prioritization}

Threat prioritization combines two complementary assessments. The severity score $s_k$ captures the technical consequences and exploitation conditions of a vulnerability, while the exploitation probability $p_k$ reflects its evolving likelihood of real-world exploitation. Keeping both quantities explicit allows threats with similar static severity to be distinguished according to current threat dynamics. Algorithm~\ref{algo:prioritization} summarizes how \system derives these two assessments from $z_k$ and combines them into the final priority ordering.

\begin{wrapfigure}{r}{0.38\columnwidth}
\vspace{0em}
\begin{minipage}{\linewidth}
\scriptsize
\begin{algorithm}[H]
  \SetAlgoLined
  \caption{Threat Prioritization
  \label{algo:prioritization}}

  \KwIn{Evidence set $\mathcal{Z}$; CVSS guidance $D_{\mathrm{CVSS}}$}
  \KwOut{Priority ordering $\pi_{\mathrm{pri}}$}

  \ForEach{$z_k\in\mathcal{Z}$}{
    \ForEach{CVSS metric $i$}{
      $\mathcal{E}_k^{(i)}
      \leftarrow
      \textsc{Select}(\mathcal{E}_k,i)$\;
      $v_k^{(i)}
      \leftarrow
      f_{\mathrm{LLM}}^{(i)}
      (\mathcal{E}_k^{(i)},D_{\mathrm{CVSS}})$\;
    }
    $s_k
    \leftarrow
    \mathcal{S}_{\mathrm{CVSS}}(\mathbf{v}_k)$\;

    $\mathcal{E}_k^{\mathrm{temp}}
    \leftarrow
    \textsc{TemporalView}(\mathcal{E}_k)$\;
    $p_k
    \leftarrow
    f_{\mathrm{LLM}}^{\mathrm{exp}}
    (z_k,\mathbf{v}_k,\mathcal{E}_k^{\mathrm{temp}})$\;

    $r_k\leftarrow(z_k,s_k,p_k)$\;
  }

  $\pi_{\mathrm{pri}}
  \leftarrow
  \textsc{Rank}(\{r_k\}_{k=1}^{M})$\;

  \Return $\pi_{\mathrm{pri}}$\;
\end{algorithm}
\end{minipage}
\vspace{-2em}
\end{wrapfigure}

\textbf{\ding{182} Evidence-grounded severity assessment.}
We represent technical severity through the CVSS vector
\smash{$\mathbf{v}_k=(v_k^{(1)},\ldots,v_k^{(d)})
\in
\mathcal{Y}^{(1)}\times\cdots\times\mathcal{Y}^{(d)}$},
where \smash{$\mathcal{Y}^{(i)}$} is the valid categorical domain of metric $i$. For each metric, \system selects a source-linked evidence view
\smash{$\mathcal{E}_k^{(i)}\subseteq\mathcal{E}_k$}
and predicts
\smash{$v_k^{(i)}
=
f_{\mathrm{LLM}}^{(i)}
(\mathcal{E}_k^{(i)},D_{\mathrm{CVSS}})$}
under the official CVSS instructions
\citep{firstCVSS31spec,firstCVSS31userguide}.
Exposure evidence informs Attack Vector, exploitation preconditions inform Attack Complexity, authentication requirements inform Privileges Required, and victim-action dependencies inform User Interaction. Confidentiality, integrity, and availability are inferred from evidence describing the consequences of successful exploitation. Detailed metric-specific workflows are provided in Appendix~\ref{app:static-analysis}.

Once all categorical metrics are determined, the official scoring function computes
$s_k=\mathcal{S}_{\mathrm{CVSS}}(\mathbf{v}_k)$,
with $s_k\in[0,10]$. The LLM thus interprets the source-linked evidence required by each CVSS dimension, while the standardized scoring procedure performs numerical aggregation.

\begin{casebox}{\bf Example 2: Evidence-grounded Attack Vector assessment}
For the threat in Example~1, \system selects the source spans ``\texttt{authenticated remote attacker}'' and ``\texttt{crafted DCERPC request}'', together with protocol evidence indicating RPC over SMB/DCERPC. When local and remote cues coexist, the model checks which access mode is supported by the documented exploitation path. The resulting evidence supports \textbf{AV=Network}. The remaining CVSS metrics are assessed independently before the official scoring procedure computes $s_k$.
\end{casebox}

\textbf{\ding{183} Temporal exploitation assessment.}
Static severity is complemented by a temporally ordered evidence view
$\mathcal{E}_k^{\mathrm{temp}}
=
\{(t_\ell,e_\ell,u_\ell)\}_{\ell=1}^{L_k}$,
where each tuple records an exploitation-related observation, its timestamp, and its source. Such observations include vulnerability disclosure, public PoC release, Exploit-DB publication, CISA KEV listing, malware observations, and reports of exploitation in the wild. \system orders these events chronologically and preserves their temporal gaps, yielding trajectories such as
$\text{disclosure}\rightarrow\text{PoC}\rightarrow\text{KEV}\rightarrow\text{in-the-wild exploitation}$.

Conditioned on the threat evidence $z_k$, its CVSS representation $\mathbf{v}_k$, and the temporal evidence $\mathcal{E}_k^{\mathrm{temp}}$, \system estimates
\smash{$p_k
=
f_{\mathrm{LLM}}^{\mathrm{exp}}
(z_k,\mathbf{v}_k,\mathcal{E}_k^{\mathrm{temp}})$},
where $p_k\in[0,1]$ denotes the probability of observed exploitation within the next 30 days, following the prediction horizon used by EPSS
\citep{jacobs2021epss,jacobs2021exploit}.
The complete set of exploitation sources is described in Appendix~\ref{app:exploit-analysis}.

\textbf{\ding{184} Joint prioritization.}
The resulting 
$r_k=(z_k,s_k,p_k)$
couples synthesized evidence with standardized severity and dynamic exploitation likelihood. Given concurrent threats, \system compares the resulting $r_k$ to produce the priority ordering $\pi_{\mathrm{pri}}$. The explicit intermediate quantities also expose the severity and exploitation evidence underlying the final ranking.

\subsection{Application II: Evidence-Grounded Mitigation Recommendation}
\label{ssec:mitigation}

\begin{wrapfigure}{r}{0.46\columnwidth}
\vspace{-1.3em}
\begin{minipage}{\linewidth}
\scriptsize
\begin{algorithm}[H]
  \SetAlgoLined
  \caption{Mitigation Synthesis
  \label{algo:mitigation}}

  \KwIn{Evidence set $\{r_k\}_{k=1}^{M}$; remediation corpus $\mathcal{D}_{\mathrm{rem}}$}
  \KwOut{Mitigation plan $\Pi_{\mathrm{mit}}$}

  \ForEach{$r_k=(z_k,s_k,p_k)$}{
    \tcp{Retrieve candidate actions}
    $\mathcal{A}_k
    \leftarrow
    f_{\mathrm{LLM}}^{\mathrm{ret}}
    (z_k,\mathcal{D}_{\mathrm{rem}})$\;

    \tcp{Normalize and verify applicability}
    $\mathcal{A}_k
    \leftarrow
    \textsc{Normalize}(\mathcal{A}_k)$\;
    $\mathcal{A}_k
    \leftarrow
    \textsc{FilterApplicable}
    (\mathcal{A}_k,z_k)$\;
  }

  \tcp{Synthesize ordered mitigation actions}
  $\Pi_{\mathrm{mit}}
  \leftarrow
  f_{\mathrm{LLM}}^{\mathrm{mit}}
  (\{(r_k,\mathcal{A}_k)\}_{k=1}^{M})$\;

  \Return $\Pi_{\mathrm{mit}}$\;
\end{algorithm}
\end{minipage}
\vspace{-1.0em}
\end{wrapfigure}

Mitigation translates the synthesized threat assessment into concrete defensive actions. Algorithm~\ref{algo:mitigation} summarizes this process from remediation retrieval to the final mitigation ordering. Given $z_k$, \system retrieves candidate mitigations from an authoritative remediation corpus $\mathcal{D}_{\mathrm{rem}}$ containing vendor security bulletins, NVD references, CISA KEV notes, and MITRE ATT\&CK mitigations. This produces
$\mathcal{A}_k
=
f_{\mathrm{LLM}}^{\mathrm{ret}}
(z_k,\mathcal{D}_{\mathrm{rem}})
=
\{a_{k,1},\ldots,a_{k,J_k}\}$,
with each action linked to its supporting source. We organize the retrieved evidence into the four categories used in our evaluation: CVE-to-patch mappings, ATT\&CK-aligned mitigations, mitigation notes or workarounds, and vendor advisories.

\textbf{\ding{182} Applicability grounding.}
Candidate actions are matched against the synthesized threat context before recommendation. Product and version metadata determine patch applicability, exploitation-path evidence identifies relevant compensating controls, and temporal evidence distinguishes current guidance from superseded remediation. \system also normalizes semantically overlapping actions while retaining their supporting provenance, producing a threat-specific action set $\mathcal{A}_k$.

\textbf{\ding{183} Mitigation ordering.}
Given the prioritized $\{r_k\}$ and associated candidate actions $\{\mathcal{A}_k\}$, \system produces
$\Pi_{\mathrm{mit}}
=
f_{\mathrm{LLM}}^{\mathrm{mit}}
(\{(r_k,\mathcal{A}_k)\}_{k=1}^{M})$.
The ordering considers threat severity and exploitation likelihood together with mitigation-specific factors such as patch availability, workaround effectiveness, implementation complexity, and operational disruption. The synthesized evidence therefore connects the evidence establishing threat urgency with the evidence supporting an appropriate defensive response.

\textbf{Output.}
\system returns the priority ordering $\pi_{\mathrm{pri}}$ together with the source-linked mitigation plan $\Pi_{\mathrm{mit}}$. Both applications operate on the same threat-centric evidence: prioritization maps evidence to urgency, while mitigation maps the resulting assessment to actionable defense.
\section{Experiment}
\label{sec:expt}

We perform extensive evaluations on \system, aiming to answer the following research questions: 
\textbf{RQ$_1$:} How effective is \system in threat prioritization and mitigation recommendation compared with existing baselines?
\textbf{RQ$_2$:} How does each component of \system contribute to its overall performance?
\textbf{RQ$_3$:} How does \system perform across individual pipeline stages?
\textbf{RQ$_4$:} Can \system generalize to zero-day vulnerability analysis?

{\bf LLMs.} 
As baseline and backbone of \system, we evaluate with five representative LLMs: Gemini-3.1-Pro (\textbf{GE3.1}), GPT-5 (\textbf{G5}), LLaMA-3-70B-Instruct (\textbf{L70B}), Claude Opus 4.1 (\textbf{C4.1}), and Qwen3-30B-A3B-Instruct-2507 (\textbf{Q3}). 
We also evaluate three LLM-based cybersecurity-specialist agents: Foundation-Sec-8B-Instruct (\textbf{FS}) \citep{cisco_foundation_sec_8b_2025}, Lily-Cybersecurity-7B (\textbf{LY}) \citep{SegoLily}, and ZySec-7B (\textbf{ZY}) \citep{ZySecAI}.

{\bf Data.}
We collect raw threat incidents from high-fidelity vendor advisories (e.g., Microsoft, Cisco, Oracle), security reports (e.g., Mandiant \citep{mandiant2004}), public CTI aggregators (e.g., the KEV Catalog \citep{CISA_KEV_Catalog}), and exploit repositories (e.g., Exploit-DB \citep{exploit-db} and GitHub PoCs). Each incident is converted into structured evaluations covering four stages of \system: (i) triage and metadata enrichment grounded in authoritative repositories such as CVE/NVD \citep{cve} and MITRE ATT\&CK \citep{strom2018attack}; (ii) CVSS assessment validated against official NVD vectors; (iii) exploitation likelihood referenced from the FIRST.org EPSS dataset \citep{jacobs2021epss}; and (iv) mitigation strategies collected from NVD references, KEV mitigation notes, and vendor security updates. We standardize incidents into stage-specific tasks and use official repository assignments as reference answers when ambiguity arises. Dataset statistics and quality-control procedures are reported in Appendix~\ref{app:benchmark}.

\textbf{Baselines.}
We compare \system with representative non-LLM and LLM-based methods for vulnerability analysis and mitigation mapping. For threat prioritization, we include \textsc{CVSS-BERT}~\citep{shahid2021cvss}, which predicts CVSS vectors from vulnerability descriptions using fine-tuned BERT classifiers, and \textsc{DeepCVA}~\citep{le2021deepcva}, which jointly predicts CVSS metrics with multi-task deep learning. Both obtain severity scores using the official CVSS formula. For mitigation recommendation, we use \textsc{CVE2CAPEC}\footnote{\url{https://github.com/Galeax/CVE2CAPEC}}, a rule-based system that maps CVEs to CWEs and CAPEC attack patterns, and \textsc{V2W-BERT}~\citep{das2021v2w}, which predicts CWE categories from CVE descriptions and retrieves associated mitigations from CWE definitions.

We also include two recent LLM-based cyber analysis frameworks as end-to-end baselines. \textsc{AutoCVSS}~\citep{sanvito2025autocvss} prompts LLMs to predict CVSS vectors and derives threat rankings from the resulting severity scores. \textsc{TRIAGE}~\citep{host2025systematic} maps CVEs to MITRE ATT\&CK techniques using rule-based prompting and in-context learning; we then retrieve the corresponding official ATT\&CK mitigations for evaluation. Since these systems have their own complete pipelines, \system is evaluated against them directly rather than layered on top. CVSS- and EPSS-based rankings serve as reference signals in our evaluation and are therefore not treated as baselines.

\subsection{Main Results: Threat Prioritization and Mitigation Recommendation (RQ$_1$)}
\label{ssec:main}

\textbf{Setting.} 
We focus on the two most decision-critical outcomes in operational CTI workflows: 
(1) \textit{threat prioritization}, evaluated by ranking-based metrics (NDCG@5 and Kendall's $\tau$) against reference urgency order, and 
(2) \textit{mitigation recommendation}, evaluated by the F1 score between generated mitigation items and database-sourced references. 
We compare \system against backbone general-purpose LLMs, cybersecurity-specialist agents, and representative non-LLM baselines.
We also compare \system with recent LLM-based cyber analysis frameworks.

\begin{table*}[t]
\centering
\caption{Evaluation of threat prioritization and mitigation recommendation (mean$\pm$std across five runs). 
Ranking quality is measured by NDCG@5 and Kendall's $\tau$, and mitigation retrieval is evaluated by F1 against reference mitigation items. ``N/A'' represents the baseline is not capable for corresponding tasks. All results are the mean and standard deviation over five runs. All metrics except Kendall's $\tau$ are reported in percentage (\%).}
\label{tab:expt-main}
\resizebox{\textwidth}{!}{
\begin{tabular}{cccccccc}
\toprule
\multicolumn{2}{c}{\multirow{2.5}{*}{\textbf{Model}}} 
& \multicolumn{2}{c}{\textbf{Prioritization Correctness}} 
& \multicolumn{4}{c}{\textbf{Mitigation Retrieval}} \\
\cmidrule(lr){3-4}\cmidrule(lr){5-8}
& & {NDCG@5} 
& {Kendall's $\tau$} 
& {CVE-to-Patch} 
& {ATT\&CK Mapping} 
& {Mitigation Note} 
& {Vendor Advisory} \\
\midrule

\rowcolor{headgray}
\multicolumn{8}{c}{\textbf{Non-LLM heuristic baseline}} \\

& \textsc{CVSS-BERT} 
& 51.2 $\pm$ 0.5
& 0.361 $\pm$ 0.012
& N/A 
& N/A 
& N/A 
& N/A \\

& \textsc{DeepCVA} 
& 55.7 $\pm$ 0.4
& 0.404 $\pm$ 0.010
& N/A 
& N/A 
& N/A 
& N/A \\

& \textsc{CVE2CAPEC} 
& N/A 
& N/A 
& 29.8 $\pm$ 0.9
& N/A 
& 24.1 $\pm$ 1.5
& 26.8 $\pm$ 0.6 \\

& \textsc{V2W-BERT} 
& N/A 
& N/A 
& 33.6 $\pm$ 0.6
& N/A 
& 28.7 $\pm$ 1.1
& 30.9 $\pm$ 0.7 \\

\midrule
\rowcolor{headgray}
\multicolumn{8}{c}{\textbf{LLM-based cyber analysis framework}} \\

& \textsc{AutoCVSS} 
& 59.8 $\pm$ 1.2
& 0.447 $\pm$ 0.007
& N/A 
& N/A 
& N/A 
& N/A \\

& \textsc{TRIAGE} 
& N/A 
& N/A 
& N/A 
& 54.2 $\pm$ 1.2
& N/A 
& N/A \\

\midrule
\rowcolor{headgray}
\multicolumn{8}{c}{\textbf{Cybersecurity-specialized agent}} \\

\multirow{3}{*}{\bf \system +}

& \textbf{FS} 
& 68.5 $\pm$ 0.8 (\goodup{8.3})
& 0.553 $\pm$ 0.009 (\goodup{0.101})
& 73.4 $\pm$ 1.3 (\goodup{11.3})
& 63.5 $\pm$ 1.5 (\goodup{13.2})
& 69.3 $\pm$ 1.4 (\goodup{9.1})
& 62.5 $\pm$ 0.9 (\goodup{7.3}) \\

& \textbf{LY} 
& 69.4 $\pm$ 0.7 (\goodup{9.2})
& 0.564 $\pm$ 0.016 (\goodup{0.112})
& 71.7 $\pm$ 0.8 (\goodup{9.6})
& 62.4 $\pm$ 0.8 (\goodup{12.1})
& 68.3 $\pm$ 0.7 (\goodup{8.1})
& 61.6 $\pm$ 0.9 (\goodup{6.4}) \\

& \textbf{ZY} 
& 70.5 $\pm$ 0.6 (\goodup{10.3})
& 0.575 $\pm$ 0.016 (\goodup{0.123})
& 74.2 $\pm$ 0.8 (\goodup{12.1})
& 64.6 $\pm$ 1.3 (\goodup{14.3})
& 70.4 $\pm$ 0.8 (\goodup{10.2})
& 63.4 $\pm$ 1.4 (\goodup{8.2}) \\

\midrule
\rowcolor{headgray}
\multicolumn{8}{c}{\textbf{General-purpose LLM}} \\

\multirow{5}{*}{\bf \system +}

& \textbf{GE3.1} 
& 85.8 $\pm$ 1.3 (\goodup{24.9})
& 0.768 $\pm$ 0.017 (\goodup{0.309})
& 92.7 $\pm$ 1.4 (\goodup{30.1})
& 80.9 $\pm$ 0.5 (\goodup{29.8})
& 89.1 $\pm$ 1.0 (\goodup{28.8})
& 81.5 $\pm$ 1.0 (\goodup{26.2}) \\

& \textbf{G5} 
& 87.3 $\pm$ 0.7 (\goodup{27.1})
& 0.783 $\pm$ 0.019 (\goodup{0.331})
& 94.3 $\pm$ 0.9 (\goodup{32.2})
& 81.5 $\pm$ 1.1 (\goodup{31.2})
& 90.3 $\pm$ 1.3 (\goodup{30.1})
& 82.3 $\pm$ 0.7 (\goodup{27.1}) \\

& \textbf{L70B} 
& 76.5 $\pm$ 0.5 (\goodup{16.3})
& 0.673 $\pm$ 0.016 (\goodup{0.221})
& 82.4 $\pm$ 1.1 (\goodup{20.3})
& 74.4 $\pm$ 1.3 (\goodup{24.1})
& 81.4 $\pm$ 1.0 (\goodup{21.2})
& 73.5 $\pm$ 0.5 (\goodup{18.3}) \\

& \textbf{C4.1} 
& 83.4 $\pm$ 0.7 (\goodup{23.2})
& 0.743 $\pm$ 0.006 (\goodup{0.291})
& 90.4 $\pm$ 0.8 (\goodup{28.3})
& 80.4 $\pm$ 1.2 (\goodup{30.1})
& 88.4 $\pm$ 1.0 (\goodup{28.2})
& 80.5 $\pm$ 0.5 (\goodup{25.3}) \\

& \textbf{Q3} 
& 80.6 $\pm$ 0.4 (\goodup{20.4})
& 0.714 $\pm$ 0.014 (\goodup{0.262})
& 87.2 $\pm$ 1.1 (\goodup{25.1})
& 77.4 $\pm$ 0.8 (\goodup{27.1})
& 86.3 $\pm$ 0.4 (\goodup{26.1})
& 78.5 $\pm$ 0.4 (\goodup{23.3}) \\

\bottomrule
\end{tabular}}
\end{table*}

{\textbf{\system consistently outperforms non-LLM baselines on both prioritization and mitigation.}}
Table~\ref{tab:expt-main} shows that severity-only predictors (e.g., \textsc{CVSS-BERT}, \textsc{DeepCVA}) provide limited prioritization accuracy, as they rely solely on predicted CVSS scores without considering exploitation likelihood or contextual threat intelligence.
In contrast, \system significantly improves both NDCG@5 and Kendall's $\tau$ by jointly reasoning over severity $(s_k)$, exploitation likelihood $(p_k)$, and enriched CTI evidence.
For mitigation recommendation, heuristic approaches like \textsc{CVE2CAPEC} achieve low F1 scores due to their reliance on predefined linkage rules that fail to capture nuanced contextual relationships.
Similarly, \textsc{V2W-BERT} is limited by its single-source knowledge dependency on CWE definitions, which often provide generic rather than actionable mitigations.
\system consistently surpasses these baselines, particularly on mitigation notes and vendor advisories that require contextual reasoning beyond pattern matching.

\subsection{Ablation Study (RQ$_2$)}
\label{ssec:ablation}

\textbf{Setting.} 
To understand the contribution of individual components in \system, we conduct an ablation study by selectively removing one stage from the pipeline at a time. 
We evaluate three representative backbone models: Gemini-3.1-Pro (\textbf{GE3.1}), GPT-5 (\textbf{G5}), and LLaMA-3-70B-Instruct (\textbf{L70B}). 

For each backbone, we consider the following ablations: 
(i) \textit{w/o Triage}, which removes incident disentanglement and metadata grounding; 
(ii) \textit{w/o Static Analysis}, which bypasses structured CVSS reasoning; and 
(iii) \textit{w/o Exploitation Analysis}, which excludes exploitation likelihood modeling. 
All variants are evaluated using the same metrics as in the main experiment.

\begin{table*}[t]
\centering
\caption{Ablation results on prioritization and mitigation (mean$\pm$std across five runs).
Each block uses a different backbone model and reports the performance of the full \system pipeline and variants with one component removed.
Results are reported as mean $\pm$ standard deviation over five runs.
All metrics except Kendall's $\tau$ are reported in percentage (\%).}
\label{tab:ablation}

\resizebox{\textwidth}{!}{
\begin{tabular}{lcccccc}
\toprule
\multirow{2}{*}{\textbf{Variant}}
& \multicolumn{2}{c}{\textbf{Prioritization Correctness}}
& \multicolumn{4}{c}{\textbf{Mitigation Retrieval}} \\
\cmidrule(lr){2-3}\cmidrule(lr){4-7}
& {NDCG@5}
& {Kendall's $\tau$}
& {CVE-to-Patch}
& {ATT\&CK Mapping}
& {Mitigation Note}
& {Vendor Advisory} \\
\midrule

\rowcolor{headgray}
\multicolumn{7}{c}{\textbf{Gemini-3.1-Pro (GE3.1)}} \\

Full \system
& 85.8 $\pm$ 1.2
& 0.768 $\pm$ 0.014
& 92.7 $\pm$ 1.0
& 80.9 $\pm$ 1.0
& 89.1 $\pm$ 1.4
& 81.5 $\pm$ 1.0 \\

w/o Triage
& 78.6 $\pm$ 1.4
& 0.679 $\pm$ 0.011
& 84.6 $\pm$ 0.6
& 73.4 $\pm$ 1.0
& 80.9 $\pm$ 0.7
& 74.2 $\pm$ 0.4 \\

w/o Static Analysis
& 81.1 $\pm$ 1.2
& 0.704 $\pm$ 0.009
& 86.8 $\pm$ 1.2
& 75.5 $\pm$ 1.1
& 83.4 $\pm$ 0.6
& 76.8 $\pm$ 0.8 \\

w/o Exploit. Analysis
& 82.4 $\pm$ 1.2
& 0.718 $\pm$ 0.009
& 87.7 $\pm$ 1.3
& 76.7 $\pm$ 0.8
& 84.9 $\pm$ 0.9
& 78.1 $\pm$ 1.2 \\

\midrule

\rowcolor{headgray}
\multicolumn{7}{c}{\textbf{GPT-5 (G5)}} \\

Full \system
& 87.3 $\pm$ 0.9
& 0.783 $\pm$ 0.019
& 94.3 $\pm$ 1.6
& 81.5 $\pm$ 1.3
& 90.3 $\pm$ 0.9
& 82.3 $\pm$ 1.5 \\

w/o Triage
& 80.1 $\pm$ 0.7
& 0.693 $\pm$ 0.013
& 86.2 $\pm$ 0.9
& 74.2 $\pm$ 1.2
& 82.1 $\pm$ 1.4
& 75.1 $\pm$ 1.3 \\

w/o Static Analysis
& 82.4 $\pm$ 1.0
& 0.718 $\pm$ 0.014
& 88.3 $\pm$ 1.4
& 76.1 $\pm$ 1.4
& 84.6 $\pm$ 0.6
& 77.6 $\pm$ 1.1 \\

w/o Exploit. Analysis
& 83.8 $\pm$ 0.6
& 0.732 $\pm$ 0.011
& 89.1 $\pm$ 1.1
& 77.4 $\pm$ 1.1
& 86.1 $\pm$ 0.7
& 78.9 $\pm$ 0.8 \\

\midrule

\rowcolor{headgray}
\multicolumn{7}{c}{\textbf{LLaMA-3-70B (L70B)}} \\

Full \system
& 76.5 $\pm$ 0.7
& 0.673 $\pm$ 0.017
& 82.4 $\pm$ 0.5
& 74.4 $\pm$ 0.7
& 81.4 $\pm$ 0.5
& 73.5 $\pm$ 1.0 \\

w/o Triage
& 73.2 $\pm$ 1.3
& 0.624 $\pm$ 0.008
& 78.1 $\pm$ 0.9
& 69.1 $\pm$ 1.1
& 75.4 $\pm$ 0.8
& 68.3 $\pm$ 1.4 \\

w/o Static Analysis
& 75.4 $\pm$ 1.0
& 0.648 $\pm$ 0.008
& 80.2 $\pm$ 0.5
& 70.9 $\pm$ 1.6
& 77.6 $\pm$ 0.7
& 70.1 $\pm$ 0.5 \\

w/o Exploit. Analysis
& 76.8 $\pm$ 0.6
& 0.662 $\pm$ 0.011
& 81.1 $\pm$ 1.1
& 72.1 $\pm$ 1.2
& 78.9 $\pm$ 0.6
& 71.4 $\pm$ 1.1 \\

\bottomrule
\end{tabular}}
\end{table*}
{\textbf{All components contribute to the overall effectiveness of \system.}}
Table~\ref{tab:ablation} shows that removing any single stage leads to consistent degradation across both prioritization and mitigation metrics. 
Among the three components, triage removal causes the largest performance drop, particularly in prioritization accuracy, highlighting the importance of disentangling raw incidents and grounding them in authoritative metadata. 
Excluding static analysis primarily affects severity-sensitive ranking and patch alignment, while removing exploitation analysis has a more pronounced impact on mitigation ordering when multiple medium-severity threats compete. 
These results indicate that \system's performance gains arise from the joint interaction of its components.

\subsection{Breakdown Analysis Across Pipeline Stages (RQ$_3$)}
\label{ssec:breakdown}

We analyze \system by breaking down its performance across individual pipeline stages: CTI triage, static analysis, and exploitation forecasting. 
This analysis contextualizes the gains observed in the main experiment (RQ$_1$) and ablation study (RQ$_2$), and highlights how each stage contributes to the overall effectiveness of the system.

{\textbf{CTI triage establishes a reliable foundation for downstream reasoning.}}
Across evaluated LLM backbones, \system consistently improves incident disentanglement and metadata enrichment, especially for entangled advisories and knowledge-grounded attributes such as ATT\&CK techniques and exploitation status. 
By converting unstructured CTI into structured, aligned representations, the triage stage reduces ambiguity in downstream reasoning and mitigates error propagation to later stages.

\begin{figure*}[t]
  \centering
  \includegraphics[width=\textwidth]{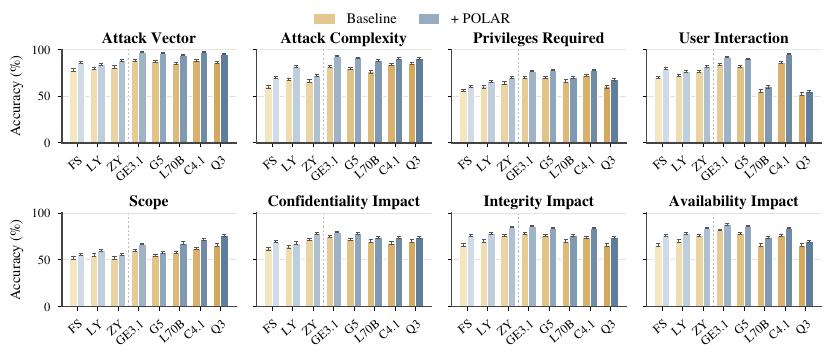}
  \caption{Static analysis performance (per-metric accuracy) across baseline models and \system.}
  \label{fig:cvss-acc-wide}
\end{figure*}

{\textbf{Static analysis benefits from evidence-grounded reasoning.}}
Figure~\ref{fig:cvss-acc-wide} shows that \system improves classification accuracy across all eight CVSS base metrics, with particularly notable gains on context-dependent dimensions such as \textit{Attack Vector} and \textit{Attack Complexity}. 
These improvements indicate that combining retrieval-grounded evidence from CTI triage with structured reasoning enables more reliable interpretation of vulnerability descriptions than severity-only prediction.

In addition to per-metric accuracy, \system also reduces overall severity scoring error, as reflected by consistently lower RMSE in Figure~\ref{fig:cvss-rmse-mini}. 
This indicates that improvements are not limited to isolated CVSS fields, but translate into more reliable aggregate severity assessments that directly support prioritization decisions.

{\textbf{Exploitation analysis enhances robustness under dynamic threat conditions.}}
For exploitation forecasting, \system shows improved robustness across different exploitation trends and historical context windows. 
While baseline models can track gradual changes, \system better adapts to abrupt shifts by linking temporal signals to broader threat narratives, which is critical when new exploits or proof-of-concept releases rapidly alter risk profiles.

Overall, the breakdown analysis demonstrates that \system’s performance gains arise from the complementary interaction of its pipeline stages rather than improvements isolated to any single component. 
Detailed quantitative results, additional figures, and per-stage analyses are provided in Appendix~\ref{app:expt-triage}, Appendix~\ref{app:expt-static}, and Appendix~\ref{app:expt-exploit}.

 \subsection{Additional Experiments (RQ$_4$)}
\label{ssec:additional}

We further evaluate \system on zero-day vulnerabilities to assess its generalization beyond disclosed and well-documented threats. 
This setting focuses on vulnerabilities that lack complete metadata or official CVE annotations at the time of analysis, reflecting realistic early-stage threat assessment scenarios.  Due to space limitations, we defer detailed analyses to the appendix: 
zero-day experiments (Appendix~\ref{app:zeroday}),
breakdown analysis with quantitative results for individual pipeline stages (Appendix~\ref{app:expt-triage}, \ref{app:expt-static}, \ref{app:expt-exploit}),
and dataset statistics (Appendix~\ref{app:benchmark}).

\section{Related Work}

\textbf{Threat prioritization.}
Traditional approaches to threat prioritization primarily relied on heuristic-driven models. Statistical and probabilistic methods, including Bayesian models and attack graphs, have been used to assess risk propagation and exploit likelihood within specific infrastructures \citep{ingols2009model, poolsappasit2012dynamic}. Other approaches integrate expert knowledge with structured taxonomies, such as MITRE’s ATT\&CK framework, to map observed adversary techniques to potential system impacts \citep{strom2018attack}. However, these methods generally emphasize static attributes and predefined rules, making them less effective in rapidly evolving threat environments.

\textbf{LLMs for cybersecurity tasks.}
Surveys report broad LLM capability in code analysis, vuln detection/repair, and broader security workflows~\citep{gwak2026lade,hasanov2024application,das2025security,zhou2025large,chen2025security}. Task-specific systems show gains in penetration testing via tool use and chain-of-thought orchestration~\citep{deng2024pentestgpt,desai2026safegpt}, and in attack modeling via retrieval-augmented generation over threat knowledge bases~\citep{prapty2024using,meng2025benchmarking,meng2025uncovering}. Evaluation and deployment studies surface challenges around robustness, hallucination, and operational risk~\citep{liu2025cylens,mcgraw202423,keppler2024evaluating}. Most prior efforts, however, target isolated tasks rather than end-to-end prioritization.

\textbf{LLM agents and automation.}
LLM-as-agent paradigms combine reasoning, tools, and environment interaction for complex goals~\citep{more2026theramind,vijayaraghavan2026hopelink,lu2025empowering,meng2026small,dong2023distdet}. In security, automated attack-graph construction and reasoning trace feasible paths and preconditions~\citep{ou2005mulval,sheyner2002automated}, while multi-agent designs specialize roles across assessment, exploitation prediction, and mitigation planning~\citep{sancho2020new}. Retrieval-augmented agents and deep log-analysis models enhance detection under real-world data heterogeneity~\citep{du2017deeplog,zhang2022deeptralog,uetz2024you}. 


\section{Conclusion}
We present \system, an LLM-powered framework for synthesizing real-world cyber evidence for threat prioritization and mitigation. \system organizes heterogeneous evidence into threat-centric representations, combines severity with temporal exploitation signals for prioritization, and connects the resulting assessments to source-linked mitigation actions. Our experiments show consistent improvements across prioritization and mitigation tasks, including heterogeneous and zero-day settings. These results demonstrate the value of evidence synthesis for practical LLM-based cyber decision support.

\bibliography{reference}
\bibliographystyle{iclr2027_conference}


\appendix

\newpage
\section{AI Use Statement}

In this work, we used generative AI tools mainly to improve the readability of the manuscript through language polishing and grammar checking. We did not use generative AI tools to generate synthetic datasets (note that the ``evidence synthesis'' in our topic means ``evidence integration'', not fabricating data); develop theoretical models or conceptual frameworks (besides that this paper studies on LLM-powered system, rather than using generative AI as tools to develop those systems); formulate mathematical claims; provide ingredients for proving mathematical claims; assist in writing proofs; propose or refine research hypotheses; design or provide feedback on research methodology or experiments; implement the proposed methods; perform translation; clean or reformat datasets; conduct qualitative or thematic data analysis; or interpret experimental results. All AI-assisted language edits were reviewed by the authors. We take responsibility for the final content of this work, including any text produced or edited with the aid of generative AI.

\section{Reproducibility Statement}
We provide an anonymized implementation and experimental materials at \mycode. The complete \system workflow and algorithms are described in \S\ref{sec:method}, with experimental settings and evaluation protocols reported in \S\ref{sec:expt}. Appendix~\ref{app:method} provides implementation instructions for evidence synthesis, CVSS assessment, temporal exploitation analysis, and mitigation recommendation, while the experimental appendix documents data collection, quality control, dataset statistics, and additional evaluations. Together with the released code, these materials are intended to support reproduction of the reported results.

\section{Limitations}

Our evaluation focuses on publicly available vulnerability and cyber threat intelligence sources and therefore does not cover organization-specific private telemetry, asset inventories, or proprietary threat feeds. We study a 30-day exploitation horizon and the mitigation categories considered in our evaluation; other operational environments may use different horizons or remediation workflows. These settings define the scope of our current evaluation, while the evidence-synthesis workflow can be adapted to additional sources and downstream security tasks.

\section{Complementary Details of \system}
\label{app:method}

We provide implementation details for the evidence synthesis, threat prioritization, and mitigation components introduced in \S\ref{sec:method}.

\subsection{Real-World Cyber Evidence Synthesis}
\label{app:evi-seynthesis}

The evidence synthesis stage converts raw cyber reports into the threat-centric evidence
$z_k=(\tau_k,\mathbf{m}_k,\mathcal{E}_k)$
used throughout \system. We first separate entangled threat descriptions and then enrich each threat with source-linked evidence and standardized metadata.

\subsubsection{Threat Disentanglement}
\label{app:disentangle}

A single CTI report can mention several vulnerabilities, campaigns, or attack behaviors. We use an LLM to separate these descriptions while preserving the context associated with each threat.

\begin{promptbox}{\small Instruction: Threat Instance Disentanglement}

\textbf{System role.}
You are a cybersecurity analyst processing cyber threat intelligence reports. Separate the input into distinct threat instances.

\vspace{3pt}
\textbf{Requirements.}
\begin{itemize}
    \item Identify explicit threat indicators such as CVE IDs, affected products, malware, infrastructure, attack techniques, and timestamps.
    \item Create one threat instance for each distinct vulnerability, malware activity, or campaign described in the report.
    \item Preserve the local context of each instance, including the affected component, exploitation behavior, required access, deployment conditions, and relevant source text.
    \item Do not merge two threats only because they appear in the same report.
    \item Do not add facts that are absent from the input.
\end{itemize}

\textbf{Input.}

\texttt{\{RAW\_INCIDENT\_TEXT\}}

\vspace{3pt}
\textbf{Output.}

Return a JSON array. Each entry contains:
\begin{itemize}
    \item \texttt{cve\_ids} if present,
    \item \texttt{vendor}, \texttt{product}, and \texttt{affected\_component},
    \item \texttt{attack\_behavior},
    \item \texttt{exploitation\_conditions},
    \item \texttt{timestamps} if present,
    \item \texttt{supporting\_spans} copied from the input.
\end{itemize}

\end{promptbox}

Each returned entry becomes a canonical threat instance $\tau_k$. We keep the supporting spans because later stages use them as evidence for severity, exploitation, and mitigation analysis.

\subsubsection{Evidence Metadata Construction and Enrichment}
\label{app:triage-db}

After disentanglement, \system builds the evidence set $\mathcal{E}_k$ and metadata $\mathbf{m}_k$ for each threat. The implementation follows four simple steps.

\textbf{Entity linking and evidence clustering.}
We first link mentions that refer to the same threat. Exact CVE matches provide the strongest key when available. When a CVE is absent, we use the affected vendor/product, component, exploitation behavior, and nearby timestamps to associate reports. Repeated descriptions from different sources are grouped under the same threat while their source information is preserved.

\textbf{Event decomposition and evidence roles.}
Long descriptions are split into evidence units that describe one relevant observation, such as a vulnerability property, PoC release, exploitation report, or mitigation statement. Each evidence unit follows the representation
$e_{k,r}=(s_{k,r},u_{k,r},t_{k,r},c_{k,r})$,
where $s_{k,r}$ is the supporting span, $u_{k,r}$ is its source, $t_{k,r}$ is its timestamp when available, and $c_{k,r}$ records its role. We use four main roles: \texttt{identity}, \texttt{severity}, \texttt{exploitation}, and \texttt{mitigation}.

\textbf{External evidence enrichment.}
We query external sources using the CVE ID and available aliases of $\tau_k$. The main sources are:
\begin{itemize}
    \item \textbf{NVD/CVE records} for canonical vulnerability descriptions, affected products and versions, disclosure dates, and references \citep{cve};
    \item \textbf{MITRE ATT\&CK} for relevant attacker techniques and behavioral mappings \citep{strom2018attack};
    \item \textbf{CISA KEV} for confirmed exploitation status and associated remediation guidance \citep{CISA_KEV_Catalog};
    \item \textbf{Exploit-DB} and other public exploit sources for PoC availability and publication dates \citep{exploit-db};
    \item \textbf{VirusTotal} for malware, URL, file, and related threat observations \citep{virustotal};
    \item \textbf{vendor advisories, CERT bulletins, CTI reports, and available telemetry} for affected configurations, exploitation updates, and remediation context.
\end{itemize}

The standardized metadata $\mathbf{m}_k$ contains fields that can be represented directly, such as the CVE ID, vendor/product, affected versions, affected component, disclosure date, and ATT\&CK mappings. Free-text observations remain in $\mathcal{E}_k$ with their provenance.

\textbf{Temporal relations and conflicting evidence.}
Timestamped evidence is normalized to a common date format and ordered by observation time. We retain simple relations such as the same threat appearing across sources and one observation occurring before another. Conflicting claims are not silently overwritten. Both claims remain in $\mathcal{E}_k$ with their sources and dates. For canonical fields such as affected versions or patch availability, we give preference to current vendor or government guidance and mark older information as superseded when the source explicitly indicates an update.

These steps produce the shared representation
$z_k=(\tau_k,\mathbf{m}_k,\mathcal{E}_k)$.
Task-specific views such as $\mathcal{E}_k^{(\mathrm{AV})}$ or
$\mathcal{E}_k^{\mathrm{temp}}$ are selected from this same evidence set.

\begin{casebox}{\bf Example: Building a Print Spooler evidence metadata}
\ul{\bf Raw report $i_j$.} A report describes both local privilege escalation and remote Print Spooler exploitation through crafted RPC requests.

\vspace{2pt}
\ul{\bf Disentanglement.} $\phi_{\mathrm{LLM}}(i_j)$ separates the report into distinct threat instances. For the remote exploitation path, $\tau_k$ keeps the Print Spooler component, remote access condition, crafted RPC behavior, and resulting code execution.

\vspace{2pt}
\ul{\bf Metadata $\mathbf{m}_k$.} \system links the instance to its CVE record, affected Windows versions, disclosure date, and relevant ATT\&CK mappings.

\vspace{2pt}
\ul{\bf Evidence $\mathcal{E}_k$.} Example evidence units include ``authenticated remote attacker'' from a vendor advisory with role \texttt{severity}, ``crafted DCERPC request'' with role \texttt{severity}, and a dated KEV observation with role \texttt{exploitation}. Together, $\tau_k$, $\mathbf{m}_k$, and $\mathcal{E}_k$ form $z_k$ for the downstream analyses.
\end{casebox}

\subsection{Threat Prioritization Details}
\label{app:threat-priori}

Threat prioritization uses two views of the same evidence metadata: metric-specific evidence for CVSS assessment and temporally ordered evidence for exploitation analysis.

\subsubsection{CVSS Metric Assessment}
\label{app:static-analysis}

For each CVSS metric $i$, \system selects
$\mathcal{E}_k^{(i)}\subseteq\mathcal{E}_k$
and provides the selected evidence, its provenance, and the official CVSS definition to the LLM. The model returns the categorical value and a short rationale pointing to the decisive evidence. Once all base metrics are assigned, we compute the final score using the official CVSS scoring procedure.

The following rules define the evidence used for each metric.

\textbf{Attack Vector (AV).}
We look for evidence describing how the vulnerable component is reached. Network services and remote requests support \texttt{Network}; communication limited to a shared local network supports \texttt{Adjacent}; execution that requires local system access supports \texttt{Local}; and physical access supports \texttt{Physical}. Protocol and deployment information in $\mathbf{m}_k$ is used to confirm the access path.

\textbf{Attack Complexity (AC).}
We check whether successful exploitation depends on conditions outside the attacker's control. Race conditions, rare system states, or required non-default configurations can support \texttt{High}. Normal attacker actions such as crafting a request or hosting a payload do not by themselves increase AC. If the documented exploitation path works under ordinary and reproducible conditions, we assign \texttt{Low}.

\textbf{Privileges Required (PR).}
We determine the minimum privileges needed before exploitation begins. Explicit terms such as ``unauthenticated'', ``authenticated user'', or ``administrator'' are mapped to \texttt{None}, \texttt{Low}, or \texttt{High}. Privileges obtained after successful exploitation are not used to determine PR.

\textbf{User Interaction (UI).}
We check whether successful exploitation requires an action from another user. Opening a malicious document or clicking a link supports \texttt{Required}. Exploitation triggered directly by attacker-controlled requests supports \texttt{None}.

\textbf{Scope (S).}
We determine whether exploitation can affect a security authority different from the vulnerable component. Evidence about sandbox escape, cross-tenant effects, privilege-domain transitions, or impact on another security boundary is used to distinguish \texttt{Changed} from \texttt{Unchanged}. When the available evidence only describes effects within the same security authority, we use \texttt{Unchanged}.

\textbf{Confidentiality, Integrity, and Availability (C/I/A).}
We assess the direct impact of successful exploitation. Information disclosure supports confidentiality impact; unauthorized modification or code execution supports integrity impact; and crashes or service disruption support availability impact. Each dimension is assigned independently according to the extent of the documented impact.

\textbf{Handling conflicting evidence.}
When sources disagree, \system checks whether the statements describe different versions, configurations, or exploitation paths. Metric assignments use the path and conditions supported by the evidence being scored and follow the official CVSS guidance when several valid paths exist. The model is instructed to cite the evidence that determines each final metric.

\begin{casebox}{\bf Example: Evidence-grounded CVSS assessment}
For the Print Spooler threat above, ``authenticated remote attacker'' and RPC/DCERPC evidence form part of $\mathcal{E}_k^{(\mathrm{AV})}$ and support \textbf{AV=Network}. The same authentication cue contributes to $\mathcal{E}_k^{(\mathrm{PR})}$ and supports \textbf{PR=Low}. If the exploitation path requires no victim action, the corresponding evidence supports \textbf{UI=None}. The remaining metrics are determined from their own evidence views before the official CVSS procedure computes $s_k$.
\end{casebox}

\subsubsection{Temporal Exploitation Assessment}
\label{app:exploit-analysis}

The exploitation component estimates how likely a vulnerability is to be observed in exploitation within the next 30 days. We construct the temporal evidence view from evidence available at the analysis time only.

\textbf{Evidence sources.}
We collect exploitation-related observations from the following sources.

\begin{itemize}
    \item \textbf{Exploit-DB:} PoC availability, exploit type, and publication date \citep{exploit-db}.
    \item \textbf{CISA KEV:} confirmed exploitation status and KEV addition date \citep{CISA_KEV_Catalog}.
    \item \textbf{VirusTotal:} first-seen malware or related file/URL observations when they can be linked to the threat \citep{virustotal}.
    \item \textbf{Vendor advisories and CERT bulletins:} dated updates concerning exploitation, PoC availability, patches, and observed campaigns \citep{cert_bulletins}.
    \item \textbf{CTI reports and available telemetry:} dated observations of scanning, malware activity, exploitation attempts, or campaign adoption.
\end{itemize}

\textbf{Timeline construction.}
For each source, we extract a short event record containing its date, event type, source, and supporting text. Dates are normalized, duplicate observations are merged while preserving all sources, and events are sorted chronologically. Evidence published after the analysis cutoff is excluded. The resulting
$\mathcal{E}_k^{\mathrm{temp}}$
may contain events such as disclosure, PoC publication, KEV listing, malware observation, or exploitation in the wild.

\textbf{Exploitation estimation.}
The LLM receives $z_k$, the CVSS representation $\mathbf{v}_k$, and the ordered temporal evidence. It is instructed to consider the order and spacing of events, evidence of weaponization, deployment exposure, patch availability, and observed attacker interest. CVSS severity alone is not treated as evidence of active exploitation.

\begin{promptbox}{\small Instruction: Temporal Exploitation Assessment}

\textbf{System role.}
Estimate the probability that the vulnerability will be observed in exploitation within the next 30 days.

\vspace{3pt}
\textbf{Input.}
\begin{itemize}
    \item threat evidence metadata $z_k$;
    \item CVSS metric vector $\mathbf{v}_k$;
    \item chronologically ordered exploitation evidence $\mathcal{E}_k^{\mathrm{temp}}$.
\end{itemize}

\textbf{Reasoning requirements.}
\begin{itemize}
    \item Use only evidence available before the analysis cutoff.
    \item Consider how quickly PoCs or weaponized exploits appeared after disclosure.
    \item Give strong weight to confirmed KEV or in-the-wild exploitation evidence.
    \item Consider whether the affected service is commonly exposed and whether patches are already available.
    \item Do not infer exploitation only from a high CVSS score.
\end{itemize}

\textbf{Output.}

Return:
\begin{itemize}
    \item \texttt{probability}: a value in $[0,1]$;
    \item \texttt{supporting\_events}: the decisive dated observations;
    \item \texttt{rationale}: a short explanation of the estimate.
\end{itemize}

\end{promptbox}

The returned probability is $p_k$. Together with $z_k$ and $s_k$, it forms the prioritization metadata
$r_k=(z_k,s_k,p_k)$
used for threat ordering.

\subsection{Mitigation Recommendation Details}
\label{app:mitigation}

The mitigation component first retrieves source-linked defensive actions and then orders the applicable actions using the threat metadata produced during prioritization.

\subsubsection{Candidate Mitigation Retrieval}

For each $z_k$, we query the remediation corpus $\mathcal{D}_{\mathrm{rem}}$ using the CVE ID, affected product and version, component, and relevant exploitation context. Official vendor guidance is checked first, followed by CISA KEV, NVD references, MITRE ATT\&CK, maintainer documentation, CERT bulletins, and trusted security advisories.

We organize retrieved material into the four categories used in our evaluation:
\begin{itemize}
    \item \textbf{CVE-to-patch mappings:} fixed versions, patch or hotfix identifiers, release dates, and applicability;
    \item \textbf{ATT\&CK-aligned mitigations:} defensive techniques and controls relevant to the observed attack path;
    \item \textbf{mitigation notes or workarounds:} temporary configuration changes, feature disabling, segmentation, isolation, or other compensating controls;
    \item \textbf{vendor advisories:} official remediation guidance and supporting references.
\end{itemize}

\begin{promptbox}{\small Instruction: Mitigation Knowledge Retrieval}

\textbf{System role.}
Retrieve authoritative mitigation actions for the supplied threat metadata.

\vspace{3pt}
\textbf{Input.}

A threat metadata $z_k$ containing the CVE ID, affected product/version, affected component, relevant exploitation evidence, and available remediation context.

\vspace{3pt}
\textbf{Source order.}
\begin{enumerate}
    \item vendor advisories and official patch bulletins;
    \item CISA KEV mitigation guidance;
    \item NVD references;
    \item MITRE ATT\&CK mitigation knowledge;
    \item official maintainer or project documentation;
    \item trusted CERT or security advisories.
\end{enumerate}

\textbf{For each candidate action, return:}
\begin{itemize}
    \item action type;
    \item recommended action;
    \item affected and fixed versions when available;
    \item source and publication/update date;
    \item applicability conditions;
    \item operational notes or limitations;
    \item whether the guidance supersedes an older action.
\end{itemize}

\textbf{Output.}

Return four groups:
\texttt{CVE-to-Patch},
\texttt{ATT\&CK Mitigation},
\texttt{Mitigation Notes/Workarounds},
and \texttt{Vendor Advisories}.

\end{promptbox}

\textbf{Normalization and applicability filtering.}
Semantically equivalent actions from different sources are merged into one candidate while all supporting sources are retained. If a newer vendor advisory explicitly replaces an older patch or workaround, the older entry is marked as superseded. Actions that do not apply to the affected product or version in $z_k$ are removed. Temporary workarounds remain available when a patch cannot yet be applied.

The resulting threat-specific candidate set is
$\mathcal{A}_k=\{a_{k,1},\ldots,a_{k,J_k}\}$.

\subsubsection{Mitigation Ordering}

The final stage receives the prioritized threat metadatas $\{r_k\}$ and their candidate actions $\{\mathcal{A}_k\}$. It uses the existing threat priority together with action-level evidence to produce the ordered mitigation plan $\Pi_{\mathrm{mit}}$. We do not introduce a separate hand-designed risk formula in this stage.

\begin{promptbox}{\small Instruction: Risk-Aware Mitigation Ordering}

\textbf{System role.}
Produce an ordered and implementable mitigation plan from the prioritized threats and their source-grounded candidate actions.

\vspace{3pt}
\textbf{Input.}
For each threat:
\begin{itemize}
    \item $r_k=(z_k,s_k,p_k)$;
    \item applicable candidate actions $\mathcal{A}_k$;
    \item available information about patch availability, workaround effectiveness, implementation effort, and operational disruption.
\end{itemize}

\textbf{Ordering rules.}
\begin{enumerate}
    \item Follow the threat priority produced by the prioritization stage.
    \item Prefer an applicable vendor patch or fixed release when it can be deployed safely.
    \item If immediate patching is not possible, place effective vendor-supported workarounds or compensating controls earlier in the plan.
    \item For threats with strong exploitation evidence or high $p_k$, prioritize actions that can reduce exposure immediately.
    \item Consider implementation complexity and operational disruption when actions provide similar protection.
    \item Do not recommend an action if its product, version, or deployment conditions do not match $z_k$.
    \item Keep the authoritative source attached to every recommendation.
\end{enumerate}

\textbf{Output.}

Return an ordered list with:
\begin{itemize}
    \item \texttt{target};
    \item \texttt{recommended\_action};
    \item \texttt{action\_type};
    \item \texttt{source};
    \item \texttt{priority\_rationale};
    \item \texttt{dependencies\_or\_operational\_notes}.
\end{itemize}

\end{promptbox}

This produces the source-linked mitigation plan $\Pi_{\mathrm{mit}}$ described in \S\ref{ssec:mitigation}. The same threat evidence therefore supports both stages of the workflow: prioritization determines which threats are urgent, and mitigation selects the actions that fit those threats.
\section{Additional Details of Experiment}
\label{app:expt-setting}

\subsection{Data Collection and Quality Control}
\label{app:benchmark}

\textbf{Raw Incident Collection.}
We collect raw threat incidents from multiple trusted sources, including vendor advisories (e.g., Microsoft, Cisco, Oracle, Adobe), security research reports (e.g., Mandiant, Unit 42, Recorded Future), public CTI aggregators (e.g., the CISA KEV catalog \citep{CISA_KEV_Catalog} and US-CERT bulletins \citep{cert_bulletins}), and exploit repositories (e.g., Exploit-DB \citep{exploit-db} and GitHub PoCs). Each incident is preserved in its original form, including textual advisories, PDF bulletins, blog posts, and structured feeds, so that the evaluation reflects the heterogeneous evidence encountered in practice. We exclude low-confidence or non-authoritative sources and require each incident to be corroborated by at least one vendor advisory or government-maintained repository. This process provides reliable and context-rich inputs for subsequent evidence synthesis and analysis.

\textbf{I: Real-World Cyber Evidence Synthesis.}
Our collection and quality-control procedure follows the evidence-profile fields defined in Appendix~\S\ref{app:triage-db}. We construct reference information for the threat instances and source-linked evidence used by \system.

\begin{enumerate}
    \item \textbf{CVE and Vulnerability Identification.}
    Each candidate threat is validated against the CVE and NVD databases. Vulnerabilities used in this evaluation require an official CVE identifier and corresponding NVD entry.

    \item \textbf{Tactics, Techniques, and Procedures (TTPs).}
    Threat instances are mapped to MITRE ATT\&CK tactics and techniques using advisory text and official ATT\&CK references. For example, a Print Spooler vulnerability may align with T1068 (\textit{Exploitation for Privilege Escalation}) and T1210 (\textit{Exploitation of Remote Services}).

    \item \textbf{Affected Systems and Vendors.}
    Vendor, product, and version mappings are verified against NVD records and vendor advisories. For each threat, we record the affected operating systems or applications, product families, versions, and relevant architectural constraints.

    \item \textbf{Exploitation and Exposure Status.}
    We cross-check exploitation evidence using the CISA KEV catalog, public exploit repositories such as Exploit-DB and Metasploit, and vendor or CERT bulletins. KEV listings and public exploit records are retained as separate source-linked observations because they capture different forms of exploitation evidence.

    \item \textbf{Disclosure and Advisory Context.}
    Each threat is enriched with disclosure information, including the initial disclosure source and the chronology of advisory publication, exploit release, patch issuance, and subsequent updates when available.
\end{enumerate}

\textbf{II: CVSS Severity Assessment.}
Reference severity labels are obtained from official NVD CVSS v3.1 vectors. For each threat instance, we retrieve the categorical vector (\texttt{AV/AC/PR/UI/S/C/I/A}) and its corresponding base score. We apply the following quality-control rules:
\begin{enumerate}
    \item If multiple CVSS entries exist because of revisions or vendor-specific scoring, we use the latest NVD-assigned vector as the reference.
    \item Ambiguous or disputed CVSS assignments are flagged and excluded from evaluation.
    \item We retain both categorical metric values (e.g., \texttt{AV:N}) and numerical severity scores for evaluation.
\end{enumerate}

\textbf{III: Temporal Exploitation Assessment.}
Reference exploitation likelihood is obtained from the FIRST.org Exploit Prediction Scoring System (EPSS). For each CVE, we retrieve the EPSS probability and percentile associated with a fixed evaluation date. EPSS estimates exploitation within the following 30 days and therefore matches the primary prediction horizon used by \system. To ensure consistent evaluation:
\begin{itemize}
    \item EPSS scores are collected using a fixed snapshot synchronized with the evaluation window.
    \item CVEs without an available EPSS score are excluded from exploitation-probability evaluation.
    \item We retain both the raw probability and percentile rank for absolute and relative comparisons.
\end{itemize}

\textbf{IV: Mitigation Recommendation.}
Reference mitigation items are collected from authoritative sources and organized into the same four categories used by \system: CVE-to-patch mappings, ATT\&CK-aligned mitigations, mitigation notes or workarounds, and vendor advisories.

\begin{enumerate}
    \item \textbf{CVE-to-Patch Mapping.}
    We retrieve patch records linked to each CVE and cross-check them against vendor advisory portals such as Microsoft MSRC, Cisco PSIRT, and Oracle CPU. Each entry records affected versions, fixed releases or patch identifiers, and publication dates. We retain vendor-confirmed or NVD-linked records and explicitly mark superseded patches.

    \item \textbf{ATT\&CK-Aligned Mitigation.}
    Defensive actions are collected from MITRE ATT\&CK's mitigation catalog and associated with the relevant ATT\&CK techniques. For example, T1210 (\textit{Exploitation of Remote Services}) may be associated with M1031 (\textit{Network Intrusion Prevention}) and M1051 (\textit{Update Software}). We retain mappings supported by official ATT\&CK references or corresponding vendor guidance.

    \item \textbf{Mitigation Notes and Workarounds.}
    We collect remediation guidance from the CISA KEV catalog and other authoritative advisories. For KEV-listed vulnerabilities, we record remediation deadlines, required actions, and interim mitigations when available. Conflicting dates or instructions are resolved using the latest vendor-issued guidance.

    \item \textbf{Vendor Advisories.}
    Vendor advisories provide the primary reference for patches, hotfixes, and configuration workarounds. We record the official advisory identifier, recommended remediation, applicable product versions, and documented operational side effects when available. Only advisories hosted on official vendor portals are used unless a secondary source is explicitly linked by the vendor.
\end{enumerate}

\textbf{Data Statistics.}
Table~\ref{tab:data-statistics} summarizes the collected incidents and corresponding evidence used across the evaluation stages.

\begin{table}[t]
\centering
\footnotesize
\caption{Dataset statistics across the evaluation stages of \system. Numbers are rounded and reported in xK format.}
\label{tab:data-statistics}
\scalebox{0.95}{
\begin{tabular}{l l r}
\toprule
\textbf{Stage} & \textbf{Detailed Item} & \textbf{Count} \\
\midrule
& Raw threat incidents collected & 12.8K \\
\midrule
\multirow{5}{*}{\textbf{I. Evidence Synthesis}}
    & CVE-mapped instances & 12.8K \\
    & MITRE ATT\&CK TTP mappings & 12.8K \\
    & Affected vendor/product metadata & 12.8K \\
    & Exploitation / exposure status annotated & 12.8K \\
    & Disclosure and advisory context enriched & 11.6K \\
\midrule
\multirow{3}{*}{\textbf{II. Severity Assessment}}
    & CVSS vectors (official NVD) & 12.8K \\
    & Metric-specific evidence sets & 12.8K \\
    & Fully scored CVSS severities & 12.8K \\
\midrule
\multirow{3}{*}{\textbf{III. Temporal Exploitation}}
    & EPSS probabilities (FIRST.org) & 12.8K \\
    & Exploit-DB / PoC references & 6.8K \\
    & Confirmed CISA KEV exploited CVEs & 0.8K \\
\midrule
\multirow{4}{*}{\textbf{IV. Mitigation Recommendation}}
    & CVE-to-patch release mappings (NVD + vendor) & 9.9K \\
    & ATT\&CK mitigation mappings & 5.2K \\
    & CISA KEV mitigation notes & 0.4K \\
    & Vendor advisories (patches / workarounds) & 8.7K \\
\bottomrule
\end{tabular}}
\end{table}

\subsection{Additional Results and Discussions}
\label{app:result}

\subsubsection{Evidence Synthesis Effectiveness}
\label{app:expt-triage}

\textbf{\underline{Sources of performance gains.}}
A closer inspection of \system outputs shows that gains in evidence synthesis mainly come from grounding raw reports in authoritative external sources. Three forms of evidence are especially useful:

\begin{enumerate}
    \item \underline{Threat-to-record linking.}
    Grounding advisory text against NVD records reduces ambiguity when several vulnerability identifiers appear together. For example, when an advisory discusses multiple Print Spooler CVEs, evidence linking allows each threat instance to retain the correct affected product and version.

    \item \underline{MITRE ATT\&CK mappings.}
    ATT\&CK grounding provides consistent behavioral context for otherwise ambiguous descriptions. For example, wording such as ``gaining higher privileges through system calls'' can be linked to T1068 (\textit{Exploitation for Privilege Escalation}) when supported by the underlying evidence.

    \item \underline{Disclosure and exploitation evidence.}
    KEV records and disclosure chronology help distinguish confirmed exploitation from less certain reports. An advisory mentioning ``public reports of exploitation,'' for example, is treated as confirmed in-the-wild exploitation only when supported by an authoritative source such as KEV or a vendor advisory.
\end{enumerate}

Together, these sources improve the quality of the threat-centric evidence profile $z_k$ before it is used for severity, exploitation, or mitigation analysis.

\textbf{\underline{Nature of mistakes.}}
Two recurring errors remain. First, some LLM backbones occasionally produce unsupported CVE identifiers when reports are long or contain several related vulnerabilities. Second, unsupported ATT\&CK mappings can appear when the model extrapolates beyond the available evidence. Both errors arise when the model fills missing information using prior knowledge rather than remaining grounded in the supplied sources.

\textbf{\underline{Detectability of unsupported outputs.}}
These errors can usually be identified with lightweight validation:
\begin{itemize}
    \item generated CVE identifiers can be checked against the CVE/NVD corpus;
    \item ATT\&CK identifiers can be validated against the official ATT\&CK taxonomy;
    \item claims of active exploitation can be checked against KEV and dated vendor advisories.
\end{itemize}
These checks complement evidence synthesis by preventing unsupported identifiers or exploitation claims from propagating to downstream assessment.

\textbf{Takeaways.}
The gains in this stage come from linking LLM reasoning to source-grounded evidence. External validation is especially useful for entity linking, ATT\&CK mapping, and exploitation status, where unconstrained generation can otherwise introduce plausible but unsupported information.

\subsubsection{Ablation Study}
\label{app:ablation}

We provide additional discussion complementing the ablation results reported in Table~\ref{tab:ablation} of \S\ref{ssec:ablation}.

\textbf{\underline{Findings.}}
Removing any component consistently reduces end-to-end performance, showing that the stages provide complementary information. Removing \textbf{Evidence Synthesis} causes substantial degradation because downstream stages lose the disentangled and source-grounded threat profile. Removing \textbf{Severity Assessment} mainly affects severity-sensitive ranking and patch alignment, while removing \textbf{Temporal Exploitation Assessment} reduces the ability to distinguish threats with similar severity but different exploitation dynamics.

\textbf{\underline{Component interactions.}}
The stages also reinforce one another. Temporal exploitation evidence is most useful when it is attached to a correctly identified threat, while CVSS assessment benefits from metric-specific evidence selected from the synthesized profile. Evidence synthesis therefore provides the shared representation on which both downstream assessments depend.

\textbf{\underline{Implications for pipeline design.}}
The modular design allows individual components to be updated without redesigning the full workflow. Improvements to threat disentanglement, evidence enrichment, CVSS interpretation, or temporal evidence collection can be incorporated while preserving the same threat-centric interface between stages.

\subsubsection{Zero-Day Analysis}
\label{app:zeroday}

We evaluate \system on zero-day vulnerabilities to study early-stage threat assessment when CVE identifiers, complete metadata, or official severity scores are unavailable at decision time. In this setting, \system relies primarily on raw incident descriptions and emerging cyber evidence.

\begin{table*}[t]
\centering
\caption{Zero-day threat prioritization and mitigation performance under partial-information settings.
All results are the mean and standard deviation over five runs.
All metrics except Kendall's $\tau$ are reported in percentage (\%).}
\label{tab:zeroday}

\resizebox{\textwidth}{!}{
\begin{tabular}{lcccc}
\toprule
\textbf{Method}
& \textbf{NDCG@5}
& \textbf{Kendall's $\tau$}
& \textbf{Mitigation Recall@5}
& \textbf{Mitigation F1} \\
\midrule

Severity-only Heuristic
& 51.2 $\pm$ 1.2
& 0.403 $\pm$ 0.019
& 46.1 $\pm$ 1.4
& 42.8 $\pm$ 0.9 \\

EPSS-inspired Ranking
& 53.8 $\pm$ 1.4
& 0.421 $\pm$ 0.015
& 48.7 $\pm$ 0.6
& 45.1 $\pm$ 1.1 \\

Backbone LLM (no framework)
& 60.3 $\pm$ 1.5
& 0.497 $\pm$ 0.013
& 55.2 $\pm$ 0.6
& 51.8 $\pm$ 1.3 \\

\system
& 71.2 $\pm$ 0.5
& 0.623 $\pm$ 0.012
& 68.4 $\pm$ 1.2
& 65.1 $\pm$ 0.5 \\

\bottomrule
\end{tabular}}
\end{table*}

\textbf{\underline{Discussion.}}
As shown in Table~\ref{tab:zeroday}, \system substantially outperforms heuristic and backbone-only baselines under partial-information conditions. Although all methods degrade relative to fully documented vulnerabilities, \system retains stronger prioritization accuracy and mitigation recall by synthesizing the evidence that remains available. Qualitative inspection also shows that \system can recover useful mitigation actions and relative threat ordering before complete vulnerability records become available.

\subsubsection{CVSS Severity Assessment Effectiveness}
\label{app:expt-static}

\textbf{\underline{Sources of \system's gains.}}
The largest gains in CVSS assessment occur when advisory text contains ambiguous or incomplete cues. \system improves these decisions by selecting evidence relevant to each CVSS dimension:

\begin{enumerate}
    \item \textit{Attack Vector (AV).}
    Advisories may describe a vulnerability as ``remotely exploitable'' without clearly identifying the required access path. Linking the affected component to protocol and deployment evidence helps resolve cases such as RPC over SMB as \texttt{AV=Network}.

    \item \textit{Attack Complexity (AC).}
    Generic LLM predictions can overlook conditions such as race requirements, unusual configurations, or deployment dependencies. Metric-specific evidence helps distinguish these cases from ordinary low-complexity exploitation.

    \item \textit{Privileges Required (PR) and User Interaction (UI).}
    Explicit evidence such as ``authenticated attacker'' or ``user must open a crafted document'' is selected from the threat profile and used directly for the corresponding CVSS decision.
\end{enumerate}

\textbf{\underline{Consistency across metrics.}}
The same evidence-grounded procedure also improves Scope and the confidentiality, integrity, and availability dimensions by aligning free-text consequences with the corresponding CVSS definitions.

\textbf{\underline{Remaining gaps.}}
Errors remain when advisories omit important technical details or when exploitation conditions differ across product versions. For example, vague descriptions such as ``may allow elevation of privilege under certain conditions'' may not contain enough information to cleanly separate privilege requirements from attack complexity. Version-specific behavior can also cause errors when the same CVE affects several platforms differently.

\textbf{Takeaways.}
Overall, the results show that selecting source-linked evidence for each CVSS metric improves both categorical metric prediction and the resulting severity score. This supports the design of \system's evidence-grounded severity assessment.

\subsubsection{Temporal Exploitation Assessment}
\label{app:expt-exploit}

\begin{figure}[t]
  \centering
  \includegraphics[width=0.55\linewidth]{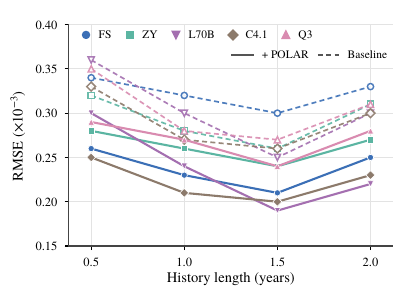}
  \caption{RMSE under varying historical context lengths for temporal exploitation assessment.}
  \label{fig:rmse-window-detailed-flat}
\end{figure}

\textbf{Experimental Setting.}
We further study temporal exploitation assessment under two conditions:
(1) different exploitation trajectories, including stable, increasing or decreasing, and abrupt-change patterns; and
(2) different lengths of historical evidence supplied to the model.
Table~\ref{tab:exploit} reports forecasting performance across exploitation trajectories, while Figure~\ref{fig:rmse-window-detailed-flat} studies the effect of historical context length. The primary \system setting estimates exploitation within 30 days; we additionally report a 90-day horizon to examine sensitivity to longer prediction windows.

\begin{table*}[ht]
\centering
\footnotesize
\renewcommand{\arraystretch}{0.9}%
\setlength{\tabcolsep}{15pt}%
\caption{Temporal exploitation forecasting by trend type using a one-year exploitation history to predict the next 30 and 90 days.
Each cell reports \system-enhanced results together with the change relative to the corresponding backbone model.
{\color{goodgreen}$\downarrow$} indicates reduced RMSE and {\color{goodgreen}$\uparrow$} indicates improved direction accuracy (DirAcc), which measures whether the exploitation trend is increasing, decreasing, or stable.
All results are the mean and standard deviation over five runs.
Results for cybersecurity-specialized agents are provided in Table~\ref{tab:more-exploit}.}
\label{tab:exploit}

\resizebox{0.98\textwidth}{!}{%
\begin{tabular}{llcccc}
\toprule
\multicolumn{2}{c}{\multirow{2.5}{*}{\bf Model}}
& \multicolumn{2}{c}{\textbf{30 Days}}
& \multicolumn{2}{c}{\textbf{90 Days}} \\
\cmidrule(lr){3-4}\cmidrule(l){5-6}
&
& \textbf{RMSE ($\times10^{-3}$) $\downarrow$}
& \textbf{DirAcc (\%) $\uparrow$}
& \textbf{RMSE ($\times10^{-3}$) $\downarrow$}
& \textbf{DirAcc (\%) $\uparrow$} \\
\midrule

\rowcolor{groupA}
\multicolumn{6}{c}{\textbf{Monotonic trend of exploitation (increasing/decreasing)}}\\
\multirow{5}{*}{\textsc{POLAR}\,+}

& GE3.1
& 1.141 $\pm$ 0.029 (\gooddown{0.587})
& 89.6 $\pm$ 0.6 (\goodup{12.1})
& 1.426 $\pm$ 0.053 (\gooddown{0.479})
& 82.1 $\pm$ 1.3 (\goodup{9.0}) \\

& G5
& 1.389 $\pm$ 0.049 (\gooddown{0.341})
& 85.2 $\pm$ 1.0 (\goodup{7.8})
& 1.738 $\pm$ 0.055 (\gooddown{0.279})
& 78.1 $\pm$ 1.5 (\goodup{5.7}) \\

& L70B
& 1.622 $\pm$ 0.057 (\gooddown{0.108})
& 81.4 $\pm$ 1.0 (\goodup{6.1})
& 2.029 $\pm$ 0.044 (\gooddown{0.089})
& 74.7 $\pm$ 1.4 (\goodup{4.5}) \\

& C4.1
& 1.098 $\pm$ 0.053 (\gooddown{0.629})
& 90.4 $\pm$ 0.6 (\goodup{13.9})
& 1.373 $\pm$ 0.047 (\gooddown{0.515})
& 82.9 $\pm$ 1.2 (\goodup{10.2}) \\

& Q3
& 1.532 $\pm$ 0.035 (\gooddown{0.198})
& 82.6 $\pm$ 1.2 (\goodup{6.4})
& 1.916 $\pm$ 0.071 (\gooddown{0.162})
& 75.9 $\pm$ 0.8 (\goodup{4.7}) \\

\rowcolor{groupB}
\multicolumn{6}{c}{\textbf{Stable trend of exploitation (no significant change)}}\\
\multirow{5}{*}{\textsc{POLAR}\,+}

& GE3.1
& 0.167 $\pm$ 0.006 (\gooddown{0.098})
& 48.7 $\pm$ 0.5 (\goodup{8.9})
& 0.210 $\pm$ 0.010 (\gooddown{0.080})
& 42.2 $\pm$ 1.1 (\goodup{6.9}) \\

& G5
& 0.215 $\pm$ 0.007 (\gooddown{0.051})
& 43.8 $\pm$ 0.4 (\goodup{5.9})
& 0.269 $\pm$ 0.007 (\gooddown{0.042})
& 37.9 $\pm$ 1.3 (\goodup{4.6}) \\

& L70B
& 0.247 $\pm$ 0.009 (\gooddown{0.019})
& 41.7 $\pm$ 0.9 (\goodup{4.0})
& 0.309 $\pm$ 0.015 (\gooddown{0.015})
& 36.1 $\pm$ 0.5 (\goodup{3.1}) \\

& C4.1
& 0.163 $\pm$ 0.004 (\gooddown{0.103})
& 49.2 $\pm$ 0.9 (\goodup{9.4})
& 0.204 $\pm$ 0.005 (\gooddown{0.084})
& 42.6 $\pm$ 1.1 (\goodup{7.3}) \\

& Q3
& 0.231 $\pm$ 0.006 (\gooddown{0.035})
& 42.7 $\pm$ 0.7 (\goodup{3.4})
& 0.290 $\pm$ 0.010 (\gooddown{0.028})
& 36.9 $\pm$ 1.4 (\goodup{2.6}) \\

\rowcolor{groupC}
\multicolumn{6}{c}{\textbf{Abrupt change of exploitation}}\\
\multirow{5}{*}{\textsc{POLAR}\,+}

& GE3.1
& 61.473 $\pm$ 2.346 (\gooddown{40.764})
& 61.4 $\pm$ 0.7 (\goodup{14.1})
& 76.385 $\pm$ 3.019 (\gooddown{33.402})
& 53.3 $\pm$ 0.8 (\goodup{11.0}) \\

& G5
& 75.212 $\pm$ 1.873 (\gooddown{27.121})
& 54.6 $\pm$ 1.5 (\goodup{7.4})
& 93.389 $\pm$ 3.641 (\gooddown{22.219})
& 47.4 $\pm$ 1.5 (\goodup{5.8}) \\

& L70B
& 84.442 $\pm$ 1.835 (\gooddown{17.891})
& 48.9 $\pm$ 1.5 (\goodup{1.7})
& 104.828 $\pm$ 2.636 (\gooddown{14.653})
& 42.4 $\pm$ 1.2 (\goodup{1.3}) \\

& C4.1
& 59.540 $\pm$ 2.626 (\gooddown{42.792})
& 62.5 $\pm$ 1.0 (\goodup{15.2})
& 73.920 $\pm$ 2.926 (\gooddown{35.068})
& 54.2 $\pm$ 1.0 (\goodup{11.8}) \\

& Q3
& 82.821 $\pm$ 3.538 (\gooddown{19.512})
& 49.6 $\pm$ 0.7 (\goodup{2.5})
& 102.841 $\pm$ 2.616 (\gooddown{15.981})
& 43.0 $\pm$ 1.0 (\goodup{1.9}) \\

\bottomrule
\end{tabular}
}
\end{table*}

Table~\ref{tab:more-exploit} complements the results in Table~\ref{tab:exploit} for cybersecurity-specialized agents.

\begin{table}[ht]
\centering
\caption{Additional temporal exploitation forecasting results by trend type, continuing Table~\ref{tab:exploit}.
All results are the mean and standard deviation over five runs.}
\label{tab:more-exploit}

\footnotesize
\renewcommand{\arraystretch}{0.9}%
\setlength{\tabcolsep}{10pt}%

\resizebox{0.9\textwidth}{!}{%
\begin{tabular}{llcccc}
\toprule
\multicolumn{2}{c}{\multirow{2.5}{*}{\bf Model}}
& \multicolumn{2}{c}{\textbf{30 Days}}
& \multicolumn{2}{c}{\textbf{90 Days}} \\
\cmidrule(lr){3-4}\cmidrule(l){5-6}
&
& \textbf{RMSE ($\times10^{-3}$)}
& \textbf{DirAcc (\%)}
& \textbf{RMSE ($\times10^{-3}$)}
& \textbf{DirAcc (\%)} \\
\midrule

\rowcolor{groupA}
\multicolumn{6}{c}{\textbf{Monotonic trend of exploitation (increasing/decreasing)}}\\
\multirow{3}{*}{\textsc{POLAR}\,+}

& FS
& 2.385 $\pm$ 0.099 (\gooddown{0.341})
& 61.8 $\pm$ 0.6 (\goodup{0.1})
& 2.928 $\pm$ 0.115 (\gooddown{0.279})
& 56.2 $\pm$ 1.3 (\goodup{0.1}) \\

& LY
& 2.216 $\pm$ 0.100 (\gooddown{0.510})
& 63.0 $\pm$ 1.3 (\goodup{0.2})
& 2.721 $\pm$ 0.113 (\gooddown{0.417})
& 57.5 $\pm$ 0.8 (\goodup{0.2}) \\

& ZY
& 2.127 $\pm$ 0.047 (\gooddown{0.599})
& 63.5 $\pm$ 0.7 (\goodup{0.3})
& 2.612 $\pm$ 0.088 (\gooddown{0.490})
& 58.0 $\pm$ 0.9 (\goodup{0.2}) \\

\rowcolor{groupB}
\multicolumn{6}{c}{\textbf{Stable trend of exploitation}}\\
\multirow{3}{*}{\textsc{POLAR}\,+}

& FS
& 0.349 $\pm$ 0.010 (\gooddown{0.003})
& 34.1 $\pm$ 1.3 (\goodup{0.2})
& 0.437 $\pm$ 0.014 (\gooddown{0.002})
& 29.6 $\pm$ 1.0 (\goodup{0.1}) \\

& LY
& 0.337 $\pm$ 0.016 (\gooddown{0.003})
& 35.3 $\pm$ 0.9 (\goodup{0.2})
& 0.422 $\pm$ 0.014 (\gooddown{0.003})
& 30.6 $\pm$ 1.0 (\goodup{0.2}) \\

& ZY
& 0.324 $\pm$ 0.009 (\gooddown{0.004})
& 36.1 $\pm$ 0.8 (\goodup{0.3})
& 0.407 $\pm$ 0.015 (\gooddown{0.003})
& 31.3 $\pm$ 1.2 (\goodup{0.2}) \\

\rowcolor{groupC}
\multicolumn{6}{c}{\textbf{Abrupt change of exploitation}}\\
\multirow{3}{*}{\textsc{POLAR}\,+}

& FS
& 102.847 $\pm$ 2.174 (\gooddown{13.485})
& 44.2 $\pm$ 1.5 (\goodup{0.5})
& 127.693 $\pm$ 4.243 (\gooddown{11.041})
& 38.4 $\pm$ 0.9 (\goodup{0.4}) \\

& LY
& 99.264 $\pm$ 3.945 (\gooddown{17.069})
& 44.9 $\pm$ 0.6 (\goodup{0.6})
& 123.208 $\pm$ 5.278 (\gooddown{13.980})
& 38.9 $\pm$ 1.6 (\goodup{0.4}) \\

& ZY
& 97.739 $\pm$ 1.979 (\gooddown{18.594})
& 45.4 $\pm$ 0.5 (\goodup{0.6})
& 121.298 $\pm$ 5.887 (\gooddown{15.227})
& 39.4 $\pm$ 0.8 (\goodup{0.5}) \\

\bottomrule
\end{tabular}
}
\end{table}

\textbf{\underline{\system's robustness across exploitation trajectories.}}
Across the trajectories in Table~\ref{tab:exploit}, \system consistently improves exploitation estimates. Baseline models can follow gradual increases or decreases but respond less reliably to sudden changes. \system performs better in these cases by connecting recent temporal signals with the surrounding exploitation history.

\begin{figure}
  \centering
  \includegraphics[width=0.5\columnwidth]{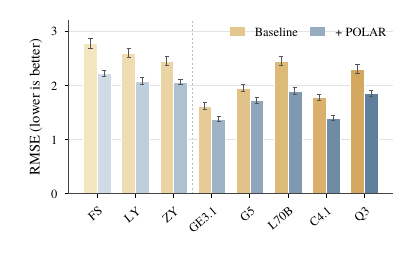}
  \caption{Overall CVSS severity scoring performance measured by RMSE.}
  \label{fig:cvss-rmse-mini}
\end{figure}

\textbf{\underline{Longer temporal context is not always beneficial.}}
Figure~\ref{fig:rmse-window-detailed-flat} shows that adding older exploitation history does not always improve prediction. Earlier observations can introduce evidence from outdated campaigns or product versions that is less relevant to the current threat state. This result suggests that temporal evidence should be selected according to the operational horizon rather than accumulated without limit. \system remains more robust than the corresponding baselines as additional context is introduced.

\textbf{\underline{Sources of \system's gains.}}
The largest improvements appear under abrupt changes in exploitation activity. In these cases, recent events such as PoC releases or new exploitation reports alter the meaning of the earlier evidence. By retaining event order and temporal context, \system adapts more quickly to these changes than models operating on less structured inputs.

\textbf{\underline{Observed gaps.}}
Although \system responds more effectively to abrupt changes, its probability estimate may still lag during the first days of a sudden exploitation surge when little precursor evidence is available.

\textbf{Takeaways.}
These results show that temporal evidence is most useful when recent observations are interpreted in the context of the broader exploitation history. Future work may further improve responsiveness through more frequent evidence updates or adaptive selection of the temporal context window.

\subsubsection{Mitigation Recommendation Effectiveness}
\label{app:expt-mitigation}

\textbf{Experimental Setting.}
We evaluate the final application from two complementary perspectives:
(1) the correctness of threat prioritization using NDCG@5 and Kendall's $\tau$, and
(2) mitigation retrieval accuracy measured by F1.
Mitigation actions are evaluated across the four categories used throughout the paper: CVE-to-patch mappings, ATT\&CK-aligned mitigations, mitigation notes, and vendor advisories.

\begin{table*}[t]
\centering
\caption{Evaluation of prioritization and mitigation recommendation across \system-enhanced cybersecurity agents and general-purpose LLM backbones.
Mitigation retrieval is measured by F1 against reference items from authoritative sources.
The reported values show the \system-enhanced result and the change (\goodup{} or \baddown{}) relative to the corresponding backbone model.}
\label{tab:expt-mitigation}
\resizebox{\textwidth}{!}{
\begin{tabular}{cccccccc}
\toprule
\multicolumn{2}{c}{\multirow{2.5}{*}{\textbf{Model}}}
& \multicolumn{2}{c}{\textbf{Prioritization Correctness}}
& \multicolumn{4}{c}{\textbf{Mitigation Retrieval}} \\
\cmidrule(lr){3-4}\cmidrule(lr){5-8}
&
& {NDCG@5}
& {Kendall's $\tau$}
& {CVE-to-Patch}
& {ATT\&CK Mapping}
& {Mitigation Note}
& {Vendor Advisory} \\
\midrule
\rowcolor{headgray}
\multicolumn{8}{c}{\textbf{Cybersecurity-specialized agent}} \\
\multirow{3}{*}{\bf \system +}
& \textbf{FS}
& 0.685 (\goodup{0.083})
& 0.553 (\goodup{0.101})
& 0.734 (\goodup{0.113})
& 0.635 (\goodup{0.132})
& 0.693 (\goodup{0.091})
& 0.625 (\goodup{0.073}) \\

& \textbf{LY}
& 0.694 (\goodup{0.092})
& 0.564 (\goodup{0.112})
& 0.717 (\goodup{0.096})
& 0.624 (\goodup{0.121})
& 0.683 (\goodup{0.081})
& 0.616 (\goodup{0.064}) \\

& \textbf{ZY}
& 0.705 (\goodup{0.103})
& 0.575 (\goodup{0.123})
& 0.742 (\goodup{0.121})
& 0.646 (\goodup{0.143})
& 0.704 (\goodup{0.102})
& 0.634 (\goodup{0.082}) \\

\midrule
\rowcolor{headgray}
\multicolumn{8}{c}{\textbf{General-purpose LLM}} \\
\multirow{5}{*}{\bf \system +}

& \textbf{GE}
& 0.793 (\goodup{0.191})
& 0.693 (\goodup{0.241})
& 0.844 (\goodup{0.223})
& 0.757 (\goodup{0.254})
& 0.845 (\goodup{0.243})
& 0.743 (\goodup{0.191}) \\

& \textbf{G5}
& 0.873 (\goodup{0.271})
& 0.783 (\goodup{0.331})
& 0.943 (\goodup{0.322})
& 0.815 (\goodup{0.312})
& 0.903 (\goodup{0.301})
& 0.823 (\goodup{0.271}) \\

& \textbf{L70B}
& 0.765 (\goodup{0.163})
& 0.673 (\goodup{0.221})
& 0.824 (\goodup{0.203})
& 0.744 (\goodup{0.241})
& 0.814 (\goodup{0.212})
& 0.735 (\goodup{0.183}) \\

& \textbf{C4.1}
& 0.834 (\goodup{0.232})
& 0.743 (\goodup{0.291})
& 0.904 (\goodup{0.283})
& 0.804 (\goodup{0.301})
& 0.884 (\goodup{0.282})
& 0.805 (\goodup{0.253}) \\

& \textbf{Q3}
& 0.806 (\goodup{0.204})
& 0.714 (\goodup{0.262})
& 0.872 (\goodup{0.251})
& 0.774 (\goodup{0.271})
& 0.863 (\goodup{0.261})
& 0.785 (\goodup{0.233}) \\

\bottomrule
\end{tabular}}
\end{table*}

\textbf{\underline{\system's effectiveness in prioritization and mitigation recommendation.}}
Results in Table~\ref{tab:expt-mitigation} show consistent gains in both threat ordering and mitigation retrieval. For prioritization, higher NDCG@5 and Kendall's $\tau$ indicate that combining severity $s_k$ with exploitation likelihood $p_k$ produces an ordering closer to the reference urgency ranking, particularly when several threats have similar static severity.

For mitigation recommendation, \system improves retrieval across all four categories. The gains are especially visible for ATT\&CK-aligned mitigations and mitigation notes, where recommendations depend on the exploitation path and surrounding threat context. These results show that the synthesized evidence profile supports both finding applicable defensive actions and connecting them to the threats that require attention.

\textbf{\underline{Sources of gains.}}
The improvement is most pronounced for mitigation categories that require evidence to be connected across sources. Baseline models can rely heavily on lexical similarity, while \system uses the threat profile to connect the affected product, exploitation behavior, and current threat evidence to corresponding defensive guidance. The same source and temporal information also helps filter outdated mitigation notes when newer vendor guidance is available.

\textbf{\underline{Remaining gaps.}}
Some errors remain for vendor-specific advisories, particularly when highly specific remediation instructions are embedded in long or inconsistently formatted documents. Sparse or conflicting severity and exploitation evidence can also affect the ordering of threats with otherwise similar profiles.

\textbf{Case Study: Competing Threats.}
In one evaluation batch, two medium-severity vulnerabilities competed for mitigation resources: one affected a widely deployed web framework with active exploit scripts, while the other affected a peripheral library with limited exposure. A severity-driven baseline assigned similar urgency to both. \system ranked the first threat higher because the synthesized evidence contained stronger exploitation signals, producing an ordering consistent with the reference threat priority.

\textbf{Practical Implications.}
The results suggest that \system is useful when many vulnerabilities compete for limited remediation capacity. Its shared evidence representation supports both threat ordering and the retrieval of applicable mitigation actions, reducing the need to repeat evidence collection separately for each decision.

\textbf{Future Directions.}
Future extensions may incorporate more structured vendor feeds, improve uncertainty handling when evidence conflicts, and model deployment constraints such as patching cost or expected downtime when constructing mitigation plans.

\end{document}